\documentclass[12pt]{article}
\usepackage{amsmath}
\usepackage{amsfonts}
\usepackage{amssymb}
\usepackage{mathptmx}
\usepackage{slashed}
\usepackage{latexsym}
\usepackage{graphicx}
\usepackage{color}

\newcommand{\newc}{\newcommand}
\newc{\be}{\begin{equation}}
\newc{\ee}{\end{equation}}
\newc{\bea}{\begin{eqnarray}}
\newc{\eea}{\end{eqnarray}}
\newc{\ol}{\overline}
\newc{\wt}{\widetilde}
\newc{\bs}{\boldsymbol}
\newc{\m}{\mathcal}
\newc{\ra}{\rightarrow}
\newc{\lra}{\leftrightarrow}
\newc{\ba}{\begin{eqnarray}}
\newc{\ea}{\end{eqnarray}}
\newc{\pa}{\partial}
\newc{\D}{\Delta}

\newc{\nn}{\nonumber}

\def\simgt{\stackrel{>}{{}_\sim}}

\def\beq{\begin{equation}}
\def\eeq{\end{equation}}
\def\bea{\begin{eqnarray}}
\def\eea{\end{eqnarray}}

\newcommand{\rr}[1]{\mathrm{#1}}

\newcommand{\pr}{\sp\prime}
\newcommand{\ov}{\overline}
\newcommand{\tn}{\tilde{n}}

\begin{document}

\begin{titlepage}

\vspace*{0.7cm}

\begin{center}
{
\bf\LARGE
E6 Models from F-theory}
\\[12mm]
James~C.~Callaghan$^{\star}$
\footnote{E-mail: \texttt{James.Callaghan@soton.ac.uk}} and 
Stephen~F.~King$^{\star}$
\footnote{E-mail: \texttt{king@soton.ac.uk}},
\\[-2mm]

\end{center}
\vspace*{0.50cm}
\centerline{$^{\star}$ \it
School of Physics and Astronomy, University of Southampton,}
\centerline{\it
SO17 1BJ Southampton, United Kingdom }
\vspace*{1.20cm}

\begin{abstract}
\noindent
F-theory is a non-perturbative formulation of type IIB superstring theory which allows for the decoupling of gravity
and for the formulation of GUT theories based on the gauge group $E_6$. 
In this paper we explore F-theory models in which the low energy supersymmetric theory
contains the particle content of three 27 dimensional representations of the underlying $E_6$ gauge group,
plus two extra right-handed neutrinos predicted from F and D flatness.
The resulting TeV scale effective theory resembles either the E6SSM or the 
NMSSM+, depending on whether an additional Abelian gauge group does or does not survive.
However there are novel features compared to both these models as follows:
(i) If the additional Abelian gauge group is unbroken then it can have a weaker gauge
coupling than in the E6SSM;
(ii) If the additional Abelian gauge group is broken then non-perturbative effects can violate
the scale invariance of the NMSSM+ leading to a generalised model;
(iii) Unification is achieved not at the field theory level but at the F-theory level 
since the gauge couplings are split by flux effects, negating the need for 
any additional doublet states which are usually required;
(iv) Proton decay is suppressed by the geometric coupling suppression of a singlet state, a mechanism peculiar to F-theory, which effectively suppresses the coupling of the exotic charge $-1/3$ colour triplet state $D$ to quarks and leptons;
(v) The $\overline{D}$ decays as a chiral leptoquark with couplings to left-handed quarks and leptons, providing characteristic and striking signatures at the LHC.
\end{abstract}

 \end{titlepage}

\thispagestyle{empty}
\vfill
\newpage

\setcounter{page}{1}


\section{Introduction}
Recently there has been a resurgence of interest in F-theory as a non-perturbative formulation of type IIB superstring theory which allows for the decoupling of gravity and for the formulation of GUT theories (first proposed in \cite{Georgi:1974sy}) based on the gauge group $E_6$ (see e.g. \cite{Heckman:2009mn} and references therein).
Although descending from a high energy $E_6$ group,
most of the models in the literature \cite{Donagi:2008ca, Beasley:2008dc, Donagi:2008kj, Beasley:2008kw, Blumenhagen:2009yv, Heckman:2010bq, Andreas:2009uf, Dudas:2009hu, Dudas:2010zb,
King:2010mq,Callaghan:2011jj,Ludeling:2011en}
focus on reproducing the minimal supersymmetric standard model (MSSM) at low energies, making it difficult to obtain an experimental link to F-theory.  In this paper we explore F-theory models in which the low energy supersymmetric theory
contains the particle content of three 27 dimensional representations of the underlying $E_6$ gauge group.  The resulting low energy models will resemble either the E6SSM \cite{King:2005jy,King:2005my,King:2007uj,Howl:2007zi,Hall:2009aj} or a generalised NMSSM+ \cite{Hall:2012mx} depending on whether an additional Abelian gauge group does or does not survive. However there are novel features compared to both these models which, if observed, would provide circumstantial evidence for F-theory.

The F-theory models considered in this paper all descend from a parent $E_8$ gauge theory \cite{Heckman:2009mn}.
A crucial question for model construction is whether a gauged $U(1)$ from the $E_8$ gauge theory can survive down to low energies, where the gauged $U(1)$ may arise from one of the Cartan generators of the non-Abelian gauge group.  A clear example of this is the case of hypercharge $U(1)_Y$, arising from SU(5) after flux breaking in many F-theory models \cite{Dudas:2010zb}.  More generally, if we begin with the case of an $E_6$ GUT group, we can break $E_6$ down to the Standard Model gauge group by turning on fluxes in $U(1)$s in the following sequence of rank preserving breakings:
\bea
E_6 & \rightarrow & SO(10)\times U(1)_{\psi} \label{SO(10)} \\
SO(10) & \rightarrow & SU(5)\times  U(1)_{\chi}  \label{SU(5)0} \\
SU(5) & \rightarrow & SU(3) \times SU(2) \times U(1)_Y \label {SM0}.
\eea
For example, the $U(1)_N$ under which the right handed neutrinos have no charge is given in terms of these $U(1)$s by,
\begin{equation}
U(1)_N =\frac{1}{4}U(1)_{\chi} +\frac{\sqrt{15}}{4}U(1)_{\psi} \label{U(1)N}.
\end{equation}
In terms of F-theory model building, the case of successive flux breaking starting with $E_6$ can be studied in the so called spectral cover formalism \cite{Dudas:2010zb}, where the model building choices relating to the particle content of the model amount to making choices about the flux breaking. 
In the F-theory models considered in this paper, there will either be a surviving gauged $U(1)_N$, or it will be broken by flux breaking at the GUT scale. 

The F-theory models with a surviving Abelian gauge group resembles the E6SSM 
\cite{King:2005jy,King:2005my,King:2007uj} which is 
a supersymmetric standard model in which precisely such an extra $U(1)_N$ gauge symmetry 
survives down to the TeV scale. However in the F-theory model the gauge coupling of the $U(1)_N$
may differ from that on the E6SSM and may be much weaker for example.
The matter spectrum is similar to that of the E6SSM, namely three 27s of E6 which ensures anomaly cancellation. This implies light exotics with the quantum numbers of Higgs doublets and colour triplets of exotic quarks, arising from three $5+\ov{5}$ representations of SU(5), plus three SU(5) singlets which are charged under $U(1)_N$. The coupling one of these singlets to $H_u H_d$ generates an effective $\mu$ term after the 
the singlet acquires a low scale vacuum expectation value (VEV). 
The F-theory version of the E6SSM does not require any additional states in order to achieve unification, unlike the E6SSM which includes an additional pair of doublet states called $H'$ and $\overline{H'}$ \cite{King:2007uj}. 
It more closely resembles the Minimal E6SSM (ME6SSM) proposed in \cite{Howl:2007zi}.
However, in the F-theory model, the gauge couplings at the GUT scale are split by flux effects, while in 
the ME6SSM, unification is achieved via an intermediate Pati-Salam gauge group.
Proton decay represents another important difference between the two models.
In the F-theory model proton decay is suppressed by the geometric coupling suppression of a singlet state, which effectively suppresses the coupling of the exotic charge $-1/3$ colour triplet state $D$ to quarks and leptons,
while in the ME6SSM all proton decay couplings are allowed but with highly suppressed coefficients.
This tends to give long lived $D$ decays in the ME6SSM, but prompt $D$ decays in the F-theory model,
with large couplings to left-handed quarks and leptons, providing characteristic and striking signatures at the LHC.

If there is no surviving extra Abelian gauge group then the F-theory model resembles the NMSSM+ which also 
involves three compete 27 dimensional families
\cite{Hall:2012mx}. However, whereas in the NMSSM+ the $U(1)_N$ is broken by an additional
sector close to the GUT scale, in the F-theory model it is simply broken by flux breaking. Another important difference is that the NMSSM+ is a scale invariant theory, involving only trilinear couplings such as the trilinear singlet couplings, while in the case of F-theory there are in addition singlet mass terms arising from non-perturbative effects, giving rise to a generalised version of the NMSSM+. The phenomenological comments in the preceding paragraph
concerning unification, proton decay and the $\overline{D}$ couplings at the LHC all apply to this case as well
where the $U(1)_N$ is broken. The main advantage of the NMSSM+ over the E6SSM is that the fine-tuning is lower due to the absence of $U(1)_N$ D-terms which would introduce a term in the Higgs potential proportional to the fourth power of the $Z'$ mass as discussed in \cite{Hall:2012mx}. The NMSSM+, involving  three compete 27 dimensional families, has lower fine-tuning than the NMSSM, which in turn has lower fine-tuning than the MSSM \cite{Hall:2012mx}, making it the lowest fine-tuned model consistent with perturbative unification.

$E_6$ based F-theory models have been discussed previously, for example, issues concerning the global resolution of $E_6$ GUTs in \cite{Kuntzler:2012bu, Cvetic:2012ts}, and the models of \cite{Chen:2010tg}.  It should be noted that here we use Abelian fluxes, whereas \cite{Chen:2010tg} uses non-Abelian fluxes.

The layout of the rest of the paper is as follows. Since many readers will not be familiar with F-theory, 
in section 2 of the paper we give a basic review of the motivation and basics of F-theory, building up to analysing the flux breaking mechanism in this way.  In section 3, the model building strategies are applied to the case of 
$E_6$ models. The D flatness conditions are considered in order to calculate the singlet VEVs of the model.  This allows us to calculate the scale at which the exotics decouple. Quark, charged lepton and neutrino masses are also discussed. In section 4, unification and proton decay are studied in the F-theory models.
In section 5, the $E_6$ based F-theory models are compared to the known E6SSM and NMSSM+ models.
Section 6 concludes the paper.

\section{A basic introduction to F-theory}
\subsection{Motivation and Basics}

A major motivation for string theory is that it provides a consistent formulation of quantum gravity, the effects of which are expected to become important at the Planck scale.  With this achievement though, comes the drawback that it is very hard to predict anything about low energy physics, due to the vast numbers of consistent solutions to the string theory equations of motion.  If, however, we follow the arguments of \cite{Heckman:2010zz} and impose the conditions of unification and decoupling on the search for realistic models, the possibilities are severely restricted.  Unification refers to the existence of a GUT structure whereby the strong, weak and electromagnetic forces are described by a single gauge group and a single coupling constant at some high energy scale.  The fact that gravity is observed to be much weaker than the other forces is linked to the term ``decoupling'', which refers to the existence of a theoretical limit where $\frac{M_{GUT}}{M_{Planck}} \rightarrow 0$.  A class of models which satisfy both the criteria of unification and decoupling are F-theory GUTs.

F-theory is a 12 dimensional, strongly coupled formulation of type IIB superstring theory.  Before considering this strongly coupled case, we can note the case of perturbative type IIB string theory, which refers to a 10 dimensional theory where 6 of the dimensions are compactified, and the string coupling constant $g_s$ (which governs how strongly the strings interact with one another) is small, $g_s < 1$.  In this perturbative regime, the particles of the Standard Model (SM) are described by excitations of open strings, whereas the graviton and gravitino are related to closed strings.  Motivated by the weakness of gravity, one could try and formulate the SM by just using open strings, the ends of which are attached to D-branes \cite{Blumenhagen:2005mu}.  In the effective field theory of such D-brane constructions, SM or GUT matter arises when strings are attached to pairs of D-brane stacks, and so these matter fields are localised along the intersections of branes called ``matter curves''.  

A problem with this perturbative setup in SU(5) GUTs, however, is that whilst we can generate a tree level Yukawa coupling for the bottom quark, we cannot for the top quark since both its chiral components live in the same GUT representation (the $10_M$ of SU(5)) and the Yukawa interaction terms with a Higgs $5_H$, namely 
$5_H 10_M10_M$, do not match up an equal number of fundamental and anti-fundamental indices \cite{Heckman:2010zz}. GUTs based on SO(10) or exceptional
groups also have problems. The spinor 16 of SO(10) cannot be realised in open
string perturbation theory, and E-type gauge groups are not possible.
This all suggests the need for extra non-perturbative ingredients. With a non-perturbative string
coupling constant $g_s \simgt 1$, exceptional gauge groups such as $E_6$, $E_7$ and $E_8$ can be realised, meaning that now, in the language of an SU(5) GUT, both the $\ov{5}_H \ov{5}_M 10_M$ and the $5_H 10_M 10_M$ couplings can be realised at tree level, due to the presence of these exceptional structures which weren't present in the case of perturbative strings.  This, therefore, leads us to the case where $g_s \simgt 1$ and F-theory.  

Formally, F-theory can be defined on a background $R^{3,1} \times X$ where $R^{3,1}$ is 4 dimensional space time, and X is a Calabi-Yau (CY) complex fourfold.  It is assumed that X is elliptically fibered with a section over a complex three-fold base, $B_3$.  What this means is that each point of the base $B_3$ is represented by a two-torus, these tori being called the fibres.  The dimensions occupied by the base are the 6 compactified dimensions of type IIB string theory, and the complex modulus of the torus fibre encodes the axion and dilaton at each point on the base \cite{Beasley:2008dc}:

\begin{equation}
\tau = C_0 +i e^{-\phi} = C_0 + \frac{i}{g_s}
\end{equation}

It is a fact that the presence of D7-branes (filling 7 spatial dimensions and 1 time dimension) affects the profile of the axio-dilaton, $\tau$ \cite{Leontaris:2012af, Katz:2011qp, Tate1975, Bershadsky:1996nh}.  As such, the reason that F-theory can be viewed as a 12 dimensional theory is that two dimensions are geometric dimensions which allow us to keep track of the variation of $\tau$ over the other ten dimensions.

In F-theory, the GUT group is realised on a 7-brane which wraps some 2 complex dimensional surface S.  We can learn a lot by studying so called 'semi-local' models, where the complications of global F-theory (as discussed in \cite{Blumenhagen:2009yv, Andreas:2009uf, Cvetic:2010rq, Cecotti:2010bp})are avoided by just looking at regions close to the GUT surface S.

\subsection{Semi-local F-theory and the Role of $E_8$}

The ideas of local F-theory focus on the submanifold S, where the GUT symmetry is localised.  We can consider intersections of the gauge brane with other 7-branes wrapping surfaces $S_i$ and supporting gauge groups $G_i$.  Along these intersections matter will reside, and so they are known as matter curves, $\Sigma_i = S \cap S_i$.  Along the matter curves , the local symmetry group is enhanced to $G_{\Sigma_i} \supset G_S \times G_i$.  We can go one step further than this and then study the intersections of matter curves at points in S.  When we have an intersection of matter curves, we induce a Yukawa coupling and there is a further enhancement of the local symmetry to $G_{\Sigma_i} \times G_{\Sigma_j} \times G_{\Sigma_k}$.  In order to study Yukawa couplings in the local setup, we can gain information by just considering the local area around the point of intersection on the surface S \cite{Leontaris:2010zd}.   

The semi-local approach to F-theory assumes that we have a parent $E_8$ gauge theory which is broken by a position dependent VEV for an adjoint Higgs field \cite{Donagi:2009ra}.  All the interactions in the theory are assumed to come from a single $E_8$ point of enhancement.  At this point, all the matter curves of the theory meet, and the local symmetry group is enhanced all the way to $E_8$.  

\subsection{An SU(5) Example and Introducing Monodromy}

As an example, we can take the GUT group on S to be SU(5).  The breaking of $E_8$ to the GUT group occurs as

\begin{equation}
E_8 \rightarrow SU(5)_{GUT} \times SU(5)_{\perp} \rightarrow SU(5)_{GUT} \times U(1)^4
\end{equation}

\noindent where the commutant of the GUT group inside $E_8$ is called the perpendicular group, and in this case is $SU(5)_{\perp}$.  The nature of the matter curves of the theory is found by decomposing the adjoint representation of $E_8$ as follows

\begin{equation}
248  \rightarrow  (24,1)+(1,24) +(10,5)+(\overline{5},10)+(\overline{5}, \overline{10})+(5,\overline{10})
\end{equation}

\noindent This equation shows us that we have twenty four singlet curves ($\theta_{ij}$), five 10 curves, and ten $\overline{5}$ curves.  The equations of these curves can be written in terms of the weights $t_i$ ($i=1,..,5$, $\sum{t_i}=0$), of the 5 representation of $SU(5)_{\perp}$ as follows

\begin{align}
         \Sigma_{10} & : t_i = 0 \notag \\
         \Sigma_{5} & : -t_i-t_j = 0, i \neq j \notag \\
         \Sigma_{1} & : \pm (t_i-t_j) = 0, i \neq j
\end{align}   
              
\noindent In general, there are non linear relations between the $t_i$ and the coefficients of the elliptic fibration, which have the effect of identifying some of the $t_i$.  The way in which the $t_i$ can be identified is determined by the 'monodromy group' \cite{Marsano:2009gv}.  As we are working in the semi-local picture, the full Calabi-Yau geometry has been decoupled, and so we must choose the monodromy group by hand.  By requiring a tree level top quark Yukawa coupling, we need at least a $Z_2$ monodromy identifying two of the weights.  This is because we need the $5_H 10_M 10_M$ coupling to be invarient under the perpendicular U(1) symmetries.  As  the top and anti-top come from the same 10 representation, they both have charge $t_i$, and the up type Higgs has charge $-t_j-t_k$, meaning that to cancel the charges we must have $2t_i -t_j -t_k = 0$.  This can only be the case for $j=k=i$, and so we must have an identification of at least two of the weights.  From now on this minimal $Z_2$ case will be assumed at all times, and we will identify $t_1 \leftrightarrow t_2$. 

In the model building section of the paper, fields will be labelled by which representations of $E_6$, SO(10) and SU(5) they transform under, and also their charge under the perpendicular U(1)s, given by the appropriate linear combination of weights, as above.

\subsection{Flux Breaking}

We can take the gauge symmetry on S to be $E_6$, $SO(10)$ or $SU(5)$ (although F-theory models with no unification group have also been studied \cite{Choi:2010nf}).  
Starting from $E_8$, there are three equivalent symmetry breaking chains that can end up with $SU(5)$, namely:

\begin{align*}
E_8 & \supset E_6 \times SU(3)_{\perp} \\
& \rightarrow  SO(10)\times U(1)_{\psi}\times SU(3)_\perp  \\
& \rightarrow  SU(5)\times  U(1)_{\chi} \times U(1)_{\psi} \times  SU(3)_\perp \\
E_8 & \supset SO(10) \times SU(4)_{\perp} \\
& \rightarrow  SU(5)\times  U(1)_{\chi} \times  SU(4)_\perp \\
E_8 & \supset SU(5) \times SU(5)_{\perp}. \\
\end{align*}

\noindent As we can see from the above breaking chains, even if we break via $E_6$ or SO(10) as the GUT group, we always end up with an $SU(5) \times U(1)^4$ structure, which subsequently breaks down to the Standard Model.  The only difference between the three pictures is which U(1)s originate from the GUT group and which originate from the perpendicular group.  Throughout this paper, we will assume that the GUT group is broken down to $SU(3) \times SU(2) \times U(1)_Y$ via flux breaking.  

There are two types of flux that can be turned on: there are fluxes in the U(1)s from the perpendicular group which preserve the chirality of complete GUT representations, and there are fluxes that can be turned on in the worldvolume of the 7-brane which break the GUT structure.  Whenever we utilise flux breaking we end up with splitting equations which tell us the net number of states in a particular representation, for example, breaking SU(5) down to the Standard Model by turning on a flux in the hypercharge direction (as discussed in \cite{Marsano:2010sq}) gives the following equations for the 10 and 5 representations of SU(5)

\ba
   10=
                           \left\{\begin{array}{ll}{\rm Rep.}&$\;\;\;\#$\\
                        n_{3,2}^1-n_{\bar 3,2}^1&:\;M_{10}\\
                       n_{\bar 3,1}^1-n_{3,1}^1& :\;M_{10}-N\\
                       n_{\bar 1,1}^1-n_{1,1}^1& :\;M_{10}+N\\
                           \end{array}\right.   &&
                           5=
                           \left\{\begin{array}{ll}{\rm Rep.}&$\;\;\;\#$\\
                        n_{3,1}^1-n_{\bar 3,1}^1&:\;M_{5}\\
                       n_{1,2}^1-n_{1,\bar 2}^1& :\;M_{5}+N\\
                           \end{array}\right. \nonumber
\ea 

\noindent We can see from these equations that the flux associated with the integer M respects the GUT structure, and so is a flux in the perpendicular U(1)s.  The flux associated with the integer N is the hypercharge flux and leads to incomplete SU(5) multiplets.  As this breaking is due to the hypercharge flux, the integer N is given by the flux dotted with the homology class of the matter curve in question.  As such, we can obtain relations between these N integers (and similar integers for different fluxes) by calculating the homology classes of the matter curves.  In order to do this, the spectral cover formalism is used, and the results are summarised in \cite{Callaghan:2011jj}.

\section{$E_6$ Models from F-theory}

We start by looking at the model of \cite{Callaghan:2011jj} (model 1), which was motivated by the fact that if we build a model based on complete 27s of $E_6$ with no matter coming from the adjoint (78) representation, we automatically take care of anomaly cancellation.  Table \ref{1} shows the model building freedom we have in choosing the M and N integers specifying the flux breaking, and how these choices determine the Standard Model particle content of the model.  Here we make the same choices for the Ms and Ns as in \cite{Callaghan:2011jj} and these choices are summarised in Table \ref{1}.  In Table \ref{1}, arbitrary numbers of singlets are allowed in the spectrum for now, so that we can calculate the restrictions on these numbers later on.  The final column of Table \ref{1} shows the low energy spectrum of the E6SSM that we want to arrive at by eliminating the required exotics from the previous column, which shows the SM particle content after flux breaking.  By comparing the final two columns of Table \ref{1}, we can see that the exotics which we wish to remove are the vector pairs $2(L+\overline{L}), Q+\overline{Q}, 2(u^c+\overline{u^c}),d^{c}+\overline{d^{c}}$ and $H_d+\overline{H_d}$.  Large masses will be generated for these fields through their coupling to SM singlet fields which acquire large VEVs.        
 
From the $E_6$ point of view, the only $E_6$ allowed trilinear term in the superpotential is $27_{t_1} 27_{t_1} 27_{t_3}$.  The vectorlike pairs which we wish to remove from the low energy particle content are those which have components in both the $27_{t_{1}}$ and $27_{t_{3}}$ multiplets.  As such, they are removed by introducing $\theta_{31}$, an $E_6$ singlet, with couplings:
\beq
\theta_{31}27_{t_1'}\overline{27_{t_3'}}= \theta_{31}Q\overline{Q}+ \theta_{31}(2u^c)(2\overline{u^c})+
 \theta_{31}d^c\overline{d^c}+ \theta_{31}(2L)(2\overline{L})+
\theta_{31}H_d\overline{H_d}.
\label{31}
\eeq

If $\theta_{31}$ gets a large VEV  these vector states get large masses as required.  The difference between this case and model 1 \cite{Callaghan:2011jj} is that in model 1, $\theta_{34}$ also gets a large VEV.  This singlet has the following couplings

\beq
\theta_{34}5_1\overline{5_2}
=\theta_{34}[3D+2H_u][3\overline{D}+3H_d]=
\theta_{34}[3(D\overline{D})]+\theta_{34}[2(H_uH_d)].
\label{34}
\eeq

In the E6SSM, these exotics are light, and so instead of getting a large VEV, this singlet now must acquire a TeV scale VEV.  It needs to be checked that the F and D-flatness constraints are satisfied, and that rapid proton decay is forbidden for the realisation of the spectrum.

\begin{table}[htdp]
\small
\begin{tabular}{|c|c|c|c|c|c|c|c|c|}
\hline
$E_6$ & $SO(10)$ & $SU(5)$  & Weight vector & $Q_N$ & $N_Y$ & $M_{U(1)}$ & SM particle content& Low energy spectrum\\
\hline
$27_{t_1'}$ & $16$ & $\overline{5}_3$ & $t_1+t_5$ & $\frac{1}{\sqrt{10}}$ & $1$ & $4$ &
$4d^c+5L$&$3d^{c}+3L$\\
\hline
$27_{t_1'}$ & $16$ & $10_M$ & $t_1$ & $\frac{1}{2\sqrt{10}}$ &$-1$ & $4$ &
$4Q+5u^c+3e^c$&$3Q+3u^{c}+3e^{c}$\\
\hline
$27_{t_1'}$ & $16$ & $\theta_{15}$ & $t_1-t_5$ & 0 & $0$ & $n_{15}$ &
$3\nu^c$&-\\
\hline
$27_{t_1'}$ & $10$ & $5_1$ & $-t_1-t_3$ & $-\frac{1}{\sqrt{10}}$ &$-1$ & $3$ &
$3D+2H_u$&$3D+2H_u$\\
\hline
$27_{t_1'}$ & $10$ & $\overline{5}_2$ & $t_1+t_4$ & $-\frac{3}{2\sqrt{10}}$ &$1$ & $3$ &
$3\overline{D}+4H_d$&$3\overline{D}+3H_d$\\
\hline
$27_{t_1'}$ & $1$ & $\theta_{14}$ & $t_1-t_4$ & $\frac{5}{2\sqrt{10}}$ &$0$ & $n_{14}$ &
$\theta_{14}$& $\theta_{14}$\\
\hline
$27_{t_3'}$ & $16$ & $\overline{5}_5$ & $t_3+t_5$ & $\frac{1}{\sqrt{10}}$ & $-1$ & $-1$ &
$\overline{d^c}+2\overline{L}$&-\\
\hline
$27_{t_3'}$ & $16$ & $10_2$ & $t_3$ & $\frac{1}{2\sqrt{10}}$ &$1$ & $-1$ &
$\overline{Q}+2\bar{u^c}$&-\\
\hline
$27_{t_3'}$ & $16$ & $\theta_{35}$ & $t_3-t_5$ & 0& $0$ & $n_{35}$ &
$-$&-\\
\hline
$27_{t_3'}$ & $10$ & $5_{H_u}$ & $-2t_1$ & $-\frac{1}{2\sqrt{10}}$ &$1$ & $0$ &
$H_u$&$H_{u}$\\
\hline
$27_{t_3'}$ & $10$ & $\overline{5}_4$ & $t_3+t_4$ & $-\frac{3}{2\sqrt{10}}$ &$-1$ & $0$ &
$\overline{H_d}$&-\\
\hline
$27_{t_3'}$ & $1$ & $\theta_{34}$ & $t_3-t_4$ & $\frac{5}{2\sqrt{10}}$ & $0$ & $n_{34}$ &
$\theta_{34}$& $\theta_{34}$ \\
\hline
- & $1$ & $\theta_{31}$ & $t_3-t_1$ & 0 & $0$ & $n_{31}$ &
$\theta_{31}$&-\\
\hline
- & $1$ & $\theta_{53}$ & $t_5-t_3$ & 0 & $0$ & $n_{53}$ &
$\theta_{53}$&-\\
\hline
- & $1$ & $\theta_{54}$ & $t_5-t_4$ & $\frac{5}{2\sqrt{10}}$ &$0$ & $n_{54}$ &
$\theta_{54}$&-\\
\hline
- & $1$ & $\theta_{45}$ & $t_4-t_5$ & $-\frac{5}{2\sqrt{10}}$ &$0$ & $n_{45}$ &
$\theta_{45}$&-\\
\hline
\end{tabular}
\caption{\small Complete $27$s of $E_6$ and their $SO(10)$ and $SU(5)$ decompositions.
The $SU(5)$ matter states decompose into SM states as
$\overline{5}\rightarrow d^c,L$ and $10\rightarrow Q,u^c,e^c$ with right-handed neutrinos
$1\rightarrow \nu^c$, while $SU(5)$ Higgs states decompose as $5\rightarrow D,H_u$ and
$\overline{5}\rightarrow \overline{D},H_d$, where $D, \overline{D}$ are exotic colour triplets and antitriplets.
We identify RH neutrinos as $\nu^c=\theta_{15}$.  Arbitrary singlets are included for giving mass to neutrinos and exotics and to ensure F- and D- flatness.}
\label{1}
\end{table}%

\subsection{$U(1)_{N}$ Charges}

The correctly normalised charge generators for $U(1)_{\psi}$ and $U(1)_{\chi}$ are 

\bea
Q_{\chi}&= & \frac{1}{2\sqrt{10}}{\rm diag}[-1,-1,-1,-1,4]\\
Q_{\psi}&= & \frac{1}{2\sqrt{6}}{\rm diag}[1,1,1,-3,0]
\label{normalised charges}
\eea

\noindent As such, from Eq. (\ref{U(1)N}), the generator for $U(1)_{N}$ is given by

\begin{equation}
Q_N = \frac{1}{2\sqrt{10}}{\rm diag}[1,1,1,-4,1] \label{QN}
\end{equation}

\noindent From this, the $U(1)_N$ charges of all the particles in the spectrum can be computed, and the results are shown in Table \ref{1}.  As required (and described in the introduction), the right handed neutrinos have zero charge under this U(1).

\subsection{Singlet VEVs and Bad Operators}

As in the previous model \cite{Callaghan:2011jj}, $\theta_{31}$ should get a string scale VEV, and as mentioned earlier $\theta_{34}$ now should get a TeV scale VEV to give mass to the exotics.  $\theta_{53}$ should get a VEV in order to generate neutrino masses (as discussed later), and in order to generate the effective $\mu$ term, $\theta_{14}$ gets a TeV scale VEV, also discussed later.  

The R-parity violating superpotential couplings $u^cd^cd^c$, $Qd^cL$, $Le^cL$, $\kappa LH_u$ as well as the dimension 5 terms in the Lagrangian corresponding to the superpotential terms $QQQL$ and $u^{c}u^{c}d^{c}e^{c}$, are forbidden by the $U(1)$ symmetries that originate in the underlying $E_{6}$.  In order to check that spontaneous symmetry breaking terms coming from SM singlet field VEVs do not allow these dangerous operators to appear, we can identify the following terms which could potentially give rise to bad operators if certain singlets acquired VEVs:  $\theta_{15}LH_{u}$, $(\theta_{31}\theta_{45}+\theta_{41}\theta_{35})10_M\overline{5_3}^2$ and $\theta_{31}\theta_{41}10_M^3\overline{5_3}$. As such, taking into account the singlet VEVs that are required, we can see that the dangerous operators do not arise provided $\theta_{15}$, $\theta_{41}$ and $\theta_{45}$ do {\it not} acquire VEVs.

However this is not sufficient to ensure the absence of baryon and lepton number violating terms because, even in the absence of these VEVs, tree level graphs can generate the dangerous operators at higher order in the singlet fields.  These issues relating to proton decay will be discussed later.  Proton decay in the context of F-theory has been previously studied, for example in \cite{Grimm:2010ez, Camara:2011nj}.

\subsection{The effective $\mu$ term}

In the E6SSM, the $\mu$ term is effectively generated when a singlet which is charged under $U(1)_N$, is coupled to $H_u H_d$ and given a TeV scale VEV.  In terms of F-theory model building, the charge of $H_u H_d$ under the perpendicular U(1) symmetries can be seen from Table \ref{1} to be $-2t_1 +t_1 +t_4 = -t_1 +t_4$.  As such, the appropriate singlet which could generate the $\mu$ term is $\theta_{14}$.  However, generating the $\mu$ term by this method requires the feature that not all the singlets will now be in the massless spectrum.  If we wanted to avoid this (the details will be discussed in more detail in the D and F-Flatness sections), we could try and generate the $\mu$ term non perturbatively, as in \cite{Callaghan:2011jj}, where non perturbative effects which break the perpendicular U(1) symmetries generate an explicit $\mu$ term which can naturally be at the electroweak scale.  However, as $H_u H_d$ is charged under $U(1)_N$, this method can't be utilised in the E6SSM, and so we must have a $\theta_{14}$ singlet in the spectrum which will acquire an electroweak scale VEV. 

\subsection{D-flatness}

In the model under consideration we assume the SUSY breaking soft masses are such that only the SM singlet fields acquire very large VEVs.  To determine them we consider
the $F$- and $D$-flatness conditions. Taking account of the  $Z_2$ monodromy, $t_1\leftrightarrow t _2$  the $D$-flatness conditions are of the form given in Eq. (\ref{Dflat}) where there are three $U_{A}(1)$s with charges given in Eq.~(\ref{charges}). We wish to show that the D-flatness conditions are satisfied by the massless fields $\theta_{31},\; \theta_{53}$ needed to give mass to exotics and, as to generate  viable neutrino masses.  Even though $\theta_{34}$ and $\theta_{14}$ get VEVs, these VEVs will be at the TeV scale whereas all the other VEVs are at the string scale.  As such, the VEV for $\theta_{34}$ and $\theta_{14}$ will be ignored in the following calculations.

\noindent The D-flatness condition for $U_{A}(1)$ is

\begin{align}
\sum_{j} Q^A_{ij} (\left|\left\langle \theta_{ij} \right\rangle \right|^2 -\left|\left\langle \theta_{ji} \right\rangle \right|^2 ) & = - \frac{Tr Q^A}{192 \pi^2} g_s^2 M_S^2 \notag \\
& = - X Tr Q^A
\label{Dflat}
\end{align}

\noindent This condition must be checked for all the U(1)s, the charge generators of which are given by

\bea
Q_{\chi}&\propto &{\rm diag}[-1,-1,-1,-1,4]\\
Q_{\psi}&\propto &{\rm diag}[1,1,1,-3,0]\\
Q_{\perp}&\propto &{\rm diag}[1,1,-2,0,0]
\label{charges}
\eea

In a general basis, $Q = {\rm diag}[t_1, t_2, t_3, t_4, t_5]$, and with just $\theta_{31}$ and $\theta_{53}$ acquiring VEVs, Eq. (\ref{Dflat}) can be written
\begin{equation}
(t_5 - t_3)|\theta_{53}|^2+(t_3 - t_1)|\theta_{31}|^2 = -X TrQ^A
\end{equation}

\noindent The trace on the right hand side of Eq. \ref{Dflat} is taken over all states, and is given by
\begin{equation}
Tr Q^A = 5 \sum n_{ij}(t_i + t_j)+10 \sum n_{k} t_k + \sum m_{ij}(t_i - t_j)
\end{equation}

For our model, this trace is computed to be
\begin{align}
Tr Q^A &= (60-n_{31}+n_{14}+n_{15}) t_1 + (n_{31}+n_{34}-n_{53}-30) t_3 + (15-n_{54}-n_{14}-n_{34}) t_4 \notag \\
&+ (15+n_{53}+n_{54}-n_{15}) t_5
\end{align}

\noindent where $n_{ij} \equiv \tilde{n}_{ij}-\tilde{n}_{ji}$ to simplify the notation, with $\tilde{n}_{ij}$ being the absolute number of $\theta_{ij}$ singlets.  Evaluating the trace for each of the U(1)s gives

\begin{align}
Tr Q_{\chi} &= 5(3-n_{15}+n_{53}+n_{54}) \\
Tr Q_{\psi} &= -15+4(n_{14}+n_{34})+n_{15}-n_{53}+3n_{54} \\
Tr Q_{\perp} &= 120 +n_{14}+n_{15}-3n_{31}-2n_{34}+2n_{53}
\end{align}

\noindent The flatness conditions with just $\theta_{31}$ and $\theta_{53}$ getting VEVs then become the three simultaneous equations

\begin{align}
5|\theta_{53}|^2 &= 5(-3+n_{15}-n_{54}-n_{53})X \label{flatness1} \\
-|\theta_{53}|^2 &= (15-n_{15}-4(n_{14}+n_{34})+n_{53}-3n_{54})X \label{flatness2} \\
2|\theta_{53}|^2 - 3|\theta_{31}|^2 &= (-120+3n_{31}-n_{14}-n_{15}+2n_{34}-2n_{53})X \label{flatness3}
\end{align}

\noindent Putting Eqs. (\ref{flatness1}) and (\ref{flatness2}) together gives the relation

\begin{equation}
n_{14}+n_{34}+n_{54}=3 \label{singlet relation 1} \\
\end{equation}

\noindent In order to cancel anomalies, we must have three complete 27s of $E_6$ and so we must have the following contraint on the absolute number of singlets

\begin{equation}
\tilde{n}_{14}+\tilde{n}_{34} = 3 \label{constraint1}
\end{equation}

\noindent If we have $\tilde{n}_{ij} \neq 0$, in general we will require that $\tilde{n}_{ji} = 0$, as otherwise we would be able to write a mass term $M \theta_{ij} \theta_{ji}$.  This is acceptable provided relations, wich will be discussed in section \ref{F section}, are satisfied.  In order to simplify the model, however, we will take the case $\tilde{n}_{ij} \neq 0 \Rightarrow \tilde{n}_{ji} = 0$, and we will take this fact to be implicit from this point onwards.  As such, Eqs. (\ref{singlet relation 1}) and (\ref{constraint1}) mean that $n_{54}=0$.  The equation for the $\theta_{53}$ VEV then becomes

\begin{equation}
|\theta_{53}|^2=(n_{15}-n_{53}-3)X \label{531}
\end{equation}   

As $\theta_{15}$ corresponds to the right handed neutrino and $\theta_{53}$ is required to give neutrino masses, both $n_{15}$ and $n_{53}$ must be positive.  Eq. (\ref{531}) then gives a lower limit on the number of right handed neutrinos in the model

\begin{equation}
\tilde{n}_{15} > 3 + \tilde{n}_{53} \label{rhn ineq}
\end{equation}
  
Due to the fact that in this model $\theta_{31}$ and $\theta_{53}$ acquire large VEVs, we require that $\tilde{n}_{31}, \, \tilde{n}_{53} \geq 1$.  Also, we must require $\tilde{n}_{34}>0$ in order to allow the exotics to get a mass via the term $\theta_{34}D\overline{D}$, and $\tilde{n}_{14}>0$ in order to generate the $\mu$ term.  We will take $\tn_{53}=1$, meaning that from Eq. (\ref{rhn ineq}), we must have $\tilde{n}_{15} > 4$.  This model will take the minimal case of 5 right handed neutrinos.  In order to satisfy Eq. (\ref{constraint1}) we choose $\tn_{14}=1$ and $\tn_{34}=2$, and we leave $\tn_{31}>0$  unspecified for now.  

\subsection{F-flatness} \label{F section}

In this model, we have taken the case where no $\theta_{ij} \theta_{ji}$ terms can be written down, so the only terms in the singlet superpotential which could generate a non zero F-term are 

\begin{equation}
W_{\theta} = \lambda_{ij} \theta_{53} \theta_{31}^i \theta_{15}^j \label{Fterm}
\end{equation}

\noindent where j corresponds to the number of right handed neutrinos and runs from 1 to 5, and the range of i represents the number of $\theta_{31}$ fields, and is yet unspecified.  Minimising the singlet potential leads to 

\begin{equation}
\frac{\partial W_{\theta}}{\partial \theta_{15}^j} = \lambda_{ij} \theta_{53} \theta_{31}^i \Rightarrow \lambda_{ij} \theta_{53} \left\langle \theta_{31}^i \right\rangle = 0 \label{Singlet potential minimum}
\end{equation}

\noindent As such, seven independent $\theta_{31}$ singlets must have zero VEVs.  We must have at least one $\theta_{31}$ which aquires a non zero VEV in order to satisfy Eq. (\ref{flatness3}), and so we choose $i=\tn_{31} = 6$.  Now we have a full singlet spectrum, consistent with F and D-flatness, where the choices we have made are given by

\begin{equation*}
\tn_{31} = 6, \, \tn_{53} = 1, \, \tn_{54} = 0, \, \tn_{14} = 1, \, \tn_{34} = 2, \, \tn_{15} = 5 
\end{equation*}


\subsubsection{Singlet mass terms}

If we were to drop the requirement that a non zero $\tn_{ij}$ means having $\tn_{ji}=0$, we could have $\theta_{ij} \theta_{ji}$ terms in the superpotential.  If, for example, neutrino masses were generated by giving a $\theta_{51}$ field a VEV the singlet superpotential would be of the form

\begin{equation}
W_{\theta} = \lambda_{ijk} \theta_{ij} \theta_{jk} \theta_{ki} + M^{ij} \theta_{15}^i \theta_{51}^j
\end{equation}

\noindent Considering the F-term for $\theta_{15}$, the relevant terms in the superpotential are

\begin{equation}
W_{\theta} = \gamma_{ij} \theta_{15}^i \theta_{53} \theta_{31}^j + M_{ik} \theta_{15}^i \theta_{51}^k \label{Mass superpotential}
\end{equation}

\noindent As such, if a $\theta_{51}$ field was to exist in the spectrum and acquire a VEV, the following relation would have to be satisfied

\begin{equation}
\frac{\partial W_{\theta}}{\partial \theta_{15}^i} = \gamma_{ij} \left\langle \theta_{31}^j \right\rangle \left\langle \theta_{53} \right\rangle + M_{ij} \left\langle \theta_{51}^j \right\rangle = 0 \label{Mass conditions0}
\end{equation}

\noindent Similarly, due to the fact that $\theta_{14}$ gets a TeV scale VEV to generate the $\mu$ term, and $\theta_{34}$ acquires a TeV VEV to give masses to the low scale exotics of the E6SSM, the presence of any $\theta_{43}$ fields in the spectrum would mean that we would have the analogous relation

\begin{equation}
\frac{\partial W_{\theta}}{\partial \theta_{43}^i} = \gamma_{ij} \left\langle \theta_{31}^j \right\rangle \left\langle \theta_{14} \right\rangle + M_{ij} \left\langle \theta_{34}^j \right\rangle = 0 \label{Mass conditions}
\end{equation}

\noindent As such, if we weren't to impose that $\theta_{ij} \neq 0 \Rightarrow \theta_{ji} =0$, the model would be consistent with F-flatness provided relations of the type in Eqs. (\ref{Mass conditions0}, \ref{Mass conditions}) were satisfied.  In our model, we take the simplest case where we don't have equations of this type. 

\subsection{Calculating the singlet VEVs}

Now we have a full spectrum for the model, we can calculate the singlet VEVs, giving us information about the scale at which the exotics decouple, neutrino masses etc.  From the D-flatness relations, we have

\begin{align}
|\theta_{53}|^2 &=(\tn_{15}-\tn_{53}-3)X  \label{flatness21} \\
3 |\theta_{31}|^2 &= 114 +3(\tn_{15}-\tn_{31})-2\tn_{34}+\tn_{14} \label{flatness22}
\end{align}

\noindent Putting the number for the singlet spectrum into these equations gives 

\begin{align}
|\theta_{53}|^2 &= X \\
|\theta_{31}|^2 &= \frac{118}{3}X 
\end{align}

\noindent where $X=\frac{g_s^2 M_S^2}{192 \pi^2}$

\subsection{Quark, charged lepton and exotic masses}

\begin{figure}[!t]
\centering
\includegraphics[scale=.7,angle=0]{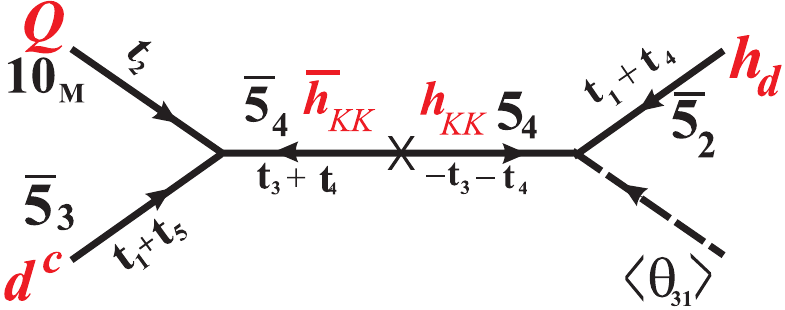}
\caption{\small{Tree-level diagram contributing to the bottom mass.}
} \label{botm}
\end{figure} 

From Table \ref{1}, we can see that the up quark mass matrix (and the Dirac neutrino mass matrix) will originate from the $27_{t_1}27_{t_1}27_{t_3}$ E6 coupling.  These matrices are rank one in the absence of flux, but as demonstrated in \cite{Aparicio:2011jx}, the rank can be increased by including non perturbative effects \cite{Cecotti:2009zf}.  The down quark and charged lepton mass matrices arise from the non-renormalisable couplings originating at the E6 level from $\theta_{31}27_{t_1}27_{t_1}27_{t_1}/M$.  Figure \ref{botm} shows the tree-level diagram for the bottom mass, involving the exchange of a massive vectorlike pair.  The origin of the difference in magnitude of the top and bottom quark masses can be explained by the fact that the $\theta_{31}$ VEV is of the same order as the messenger mass, M, leading to a mild suppression of the down quark Yukawas relative to the up quark couplings.

The terms in the superpotential which are responsible for generating the $\mu$ term and the exotic masses are

\begin{equation}
W \sim \lambda_{ij} \theta_{14} H_{d i} H_{u j} + \kappa_{ijk} \theta_{34}^i \ov{D}_j D_k
\end{equation}

\noindent From Table \ref{1}, it can be seen that both of these couplings originate from the $27_{t_1}27_{t_1}27_{t_3}$ E6 coupling.

In the standard E6SSM, an approximate $Z_2$ flavour symmetry is assumed, in order to distinguish the active (third) generations of Higgs doublets from the inert (first and second) generations.  However, in this paper we don't try and solve problems with flavour, as we can always note that in the absense of flux, matrices are always rank one.  As such, we can always pick a basis where the matrix has a one in the position corresponding to the active generation and zeros elsewhere.  Also, it should be noted that from Table \ref{1}, we can see that all three generations of $H_d$ come from the $27_{t_1}$ curve, whereas the active $H_u$ comes from a different curve ($27_{t_3}$) than the inert $H_u$s ($27_{t_1}$).  As such, we could generate the up quark masses via the non-renormalisable coupling $\theta_{31}27_{t_1}27_{t_1}27_{t_1}/M$, with $H_u$ coming from the $27_{t_1}$ matter curve.  In this case, the quark masses would arise from diagrams similar to Figure \ref{botm}.  $H_u$ will now come from the from the $5_{1}$ curve, and the diagram will involve the coupling $\theta_{31} \ov{5}_{H_u} 5_1$.  However, this coupling will turn out to be forbidden under a discrete $Z_2$ symmetry which will be introduced later in order to stabilise the proton, and so quark masses won't be generated in this manner.  In any case, it wouldn't pose a problem, due to the fact that the $\theta_{31}$ VEV is of the same order as the messenger mass, M.

\subsection{Neutrino Masses}

Due to the $t_1 \leftrightarrow t_2$ monodromy, the conjugate states $\theta_{12}$ and $\theta_{21}$ are identified, and so we can write down a term $M_M \theta_{12} \theta_{21}$ in the superpotential which corresponds to a Majorana mass for the $\theta_{12}$ states.  Using the same notation as \cite{Callaghan:2011jj}, the right handed neutrinos, $\theta_{15}$, couple to the Majorana states through the term $\lambda_{RM}^{ij} \Theta_{51} \theta_{12}^i \theta_{15}^j$, where $\Theta_{51} = \frac{\theta_{53} \theta_{31}}{M}$.  As both $\theta_{53}$ and $\theta_{31}$ acuire VEVs, $\Theta_{15}$ also has a VEV.  As in \cite{Callaghan:2011jj}, we allow for an arbitrary number of $\theta_{12}$ fields (the fact that these fields carry no charge under the perpendicular U(1)s means that we can have any number of them in the spectrum without affecting flatness conditions etc.), but the difference in this paper is that now the number of $\theta_{15}$ fields is 5, not 3.  (For a reference on models with Z right handed neutrinos, see \cite{King:1999mb}).  

The method of generating masses for the light neutrinos will be a double see-saw mechanism, where the $\theta_{15}$ fields will get Majorana masses through their coupling to the Majorana states $\theta_{12}$, and then the light neutrinos will get masses via a see saw mechanism, made possible by their coupling to the right handed neutrinos $\theta_{15}$.  The relevant terms for lepton mass generation are (after the two Higgs doublets have got their VEVs):

\begin{equation}
W_{mass} = \left\langle H_d \right\rangle Y_e^{ij} \ov{e}_L^i e_R^j + \left\langle H_u \right\rangle \lambda_{LR}^{ia} \ov{\nu}_L^i \theta_{15}^a+\left\langle \Theta_{51} \right\rangle \lambda_{RM}^{a \alpha} \theta_{15}^a \theta_{12}^{\alpha} + M_M^{\alpha \beta} \theta_{12}^{\alpha} \theta_{21}^{\beta} \label{see saw W}
\end{equation}

\noindent where $\lambda_{LR}$ is a $(3 \times 5)$ matrix of couplings, $\lambda_{RM}$ is $(5 \times n)$ (where n is the number of $\theta_{12}$ states) and $M_{M}$ is an $(n \times n)$ matrix.  We can put the notation into a more familiar form by writing $M_e^{ij} \equiv \left\langle H_d \right\rangle Y_e^{ij}, \, m_{LR}^{ia} \equiv \left\langle H_u \right\rangle \lambda_{LR}^{ia}, \, M_{RM}^{a \alpha} \equiv \left\langle \Theta_{51} \right\rangle \lambda_{RM}^{a \alpha}$.  Also, for clarity, we can relabel the fields as $\theta_{15} \equiv \nu_R , \, \theta_{12} \equiv S_R$.  Eq. (\ref{see saw W}) can then be written

\begin{equation}
W_{mass} = M_e^{ij} \ov{e}_L^i e_R^j + m_{LR}^{ia} \ov{\nu}_L^i \nu_R^a+M_{RM}^{a \alpha} \nu_R^a S_R^{\alpha} + M_M^{\alpha \beta} S_R^{\alpha} S_R^{\beta} \label{see saw W2}
\end{equation}

\noindent In the basis ($\nu_L$, $\nu_R$, $S_R$), the mass matrix is, in block form

\[
 M =
 \begin{pmatrix}
  0 & m_{LR} & 0   \\
  m_{LR} & 0 & M_{RM}\\
  0 & M_{RM} & M_M
 \end{pmatrix}
\]

\noindent Applying the double see-saw mechanism, we have (in matrix notation) for the light left-handed Majorana neutrino masses \cite{King:2003jb}

\begin{equation}
m_{LL} = m_{LR} M_{RM}^{-1} M_M (M_{RM}^{T})^{-1} m_{LR}^T \label{double see saw}
\end{equation}

\section{Unification and proton decay}

\subsection{Review of F-theory unification in $SU(5)$}
In the case where a $U(1)_Y$ flux mechanism is used to break an
$SU(5)$ gauge symmetry down to the Standard
Model, there is a splitting of the gauge couplings at the unification scale~\cite{Blumenhagen:2008aw, Conlon:2009qa, Leontaris:2009wi, Leontaris:2011pu, Leontaris:2011tw}.
The splitting at $M_{GUT}$ is
\be\label{gcMU}
\begin{split}
\frac{1}{\alpha_3(M_G)}&=\frac{1}{\alpha_G}-y
\\
\frac{1}{\alpha_2(M_G)}&=\frac{1}{\alpha_G}-y+x
\\
\frac{1}{\alpha_1(M_G)}&=\frac{1}{\alpha_G}-y+\frac 35 x
\end{split}
\ee
\noindent where $x=- \frac 12 {\rm Re} S\int c_1^2({\cal L}_Y)$, $y=\frac 12 {\rm Re} S\int c_1^2({\cal L}_a) $
${\cal L}_a$ is a non-trivial line bundle  and $S=e^{-\phi}+i\,C_0$ is the
axion-dilaton field as discussed in~\cite{Blumenhagen:2008aw}.
Combining the above, the gauge couplings at $M_{GUT}$ are found to  satisfy  the  relation
\be
\frac{1}{\alpha_Y(M_{GUT})}=\frac 53 \,\frac{1}{\alpha_1(M_{GUT})}=\frac{1}{\alpha_2(M_{GUT})}+\frac 23 \frac{1}{\alpha_3(M_{GUT})}\label{SR2}
\ee

In the E6SSM, however, we have an extra $U(1)_N$ symmetry which survives down to low energies.  Accordingly, we must incorporate the $U(1)_N$ gauge coupling into the unification analysis.  In order to acomplish this, we can consider how Eq. (\ref{SR2}) is derived in \cite{Ellis:1985jn} in terms of SU(5) group theory, and then generalise the results to $E_6$ and SO(10), giving us information about $U(1)_{\psi}$ and $U(1)_{\chi}$ respectively.

Following \cite{Ellis:1985jn}, we can write the gauge kinetic functions for SU(3), SU(2) and $U(1)_Y$ embedded inside SU(5) in the form

\begin{align}
f_3 &= A + B c_{\alpha}, \, \, \alpha=(1,...,8) \label{f3} \\
f_2 &= A + B c_{\alpha}, \, \, \alpha=(21,22,23) \label{f2} \\
f_1 &= A + B c_{\alpha}, \, \, \alpha=24 \label{f1} 
\end{align}

\noindent where $\alpha$ is an index running from 1 to 24, over all the generators of SU(5), and the missing $\alpha$s are the generators belonging outside the $SU(3) \times SU(2) \times U(1)$ subgroup of SU(5).  A and B are arbitrary gauge invariant functions and the $c_{\alpha}$ coefficients are given by

\begin{equation}
d_{\alpha \beta 24} = c_{\alpha} \delta_{\alpha \beta}
\end{equation}

\noindent with the index 24 corresponding to the hypercharge generator and the group theory coefficients $d_{\alpha \beta \gamma}$ defined as

\begin{equation}
d_{\alpha \beta \gamma} = 2 Tr [\left\{T_{\alpha},T_{\beta}\right\}  T_{\gamma}] \label{ddefinition}
\end{equation}

As such, in order to calculate the three gauge kinetic functions, we just need $d_{1,1,24}$, $d_{21,21,24}$ and $d_{24,24,24}$, where the generators $T_1$, $T_{21}$ and $T_{24}$ are given in block matrix notation by

\[
 T_{1} =
 \begin{pmatrix}
  \lambda_1 /2 & 0   \\
  0 & 0
 \end{pmatrix}
\]

\[
 T_{1} =
 \begin{pmatrix}
  0 & 0   \\
  0 & \sigma_1 /2
 \end{pmatrix}
\]

\begin{equation*}
T_{24} = \frac{1}{\sqrt{15}} diag(1,1,1,-\frac{3}{2},-\frac{3}{2})
\end{equation*}

\noindent where $\lambda_1$ refers to the first Gell-Mann matrix, and $\sigma_1$ to the first Pauli matrix.  These definitions can be used trivially to calculate $c_1 = d_{1,1,24} = \frac{2}{\sqrt{15}}$, $c_{21} = d_{21,21,24} = -\frac{3}{\sqrt{15}}$ and $c_{24} = d_{24,24,24} = -\frac{1}{\sqrt{15}}$, which can be put together with Eqs. (\ref{f3}, \ref{f2}, \ref{f1}) giving (after a redefinition of the arbitrary function B)

\begin{align}
f_3 &= A + 2B  \label{f3p} \\
f_2 &= A - 3B  \label{f2p} \\
f_1 &= A - B  \label{f1p} 
\end{align}

\noindent The gauge couplings at the unification scale are then related by \cite{Ellis:1985jn}

\begin{equation}
\alpha_G = \alpha_3 (M_G) f_3 = \alpha_2 (M_G) f_2 = \alpha_1 (M_G) f_1 = \frac{5}{3} \alpha_Y (M_G) f_1 \label{coupling unification}
\end{equation}

\noindent Combining this equation with Eqs. (\ref{f3p}, \ref{f2p}, \ref{f1p}) gives the following constraint on the gauge kinetic functions

\begin{equation}
f_3 + \frac{3}{2} f_2 = \frac{5}{2} f_1
\end{equation}

\noindent which, when combined with the relations $f_i = \frac{\alpha_G}{\alpha_i (M_G)}$, leads to Eq. (\ref{SR2}).  Comparing this picture with Eq. (\ref{gcMU}), we have the following equations relating x and y to A and B

\begin{equation}
x = -\frac{5B}{\alpha_G}, \, \, y=\frac{1-A-2B}{\alpha_G} \label{xyAB}
\end{equation}

\subsection{The $E_6$ and SO(10) cases}

We can generalise the SU(5) argument to the breaking patterns

\bea
E_6 & \rightarrow & SO(10)\times U(1)_{\psi} \nn \\
SO(10) & \rightarrow & SU(5)\times  U(1)_{\chi} \nn 
\eea

\noindent in order to learn about the $U(1)_N$ gauge coupling $U(1)_N =\frac{1}{4}U(1)_{\chi} +\frac{\sqrt{15}}{4}U(1)_{\psi}$.  For the $E_6$ case, the generalisation is the set of equations

\begin{align}
\alpha_6 &= \alpha_{10} f_{10} = \alpha_{\psi} f_{\psi} \label{E6couplings} \\
f_{10} &= A \pr + B \pr c_{\alpha}, \, \, \alpha=(1,...,45) \label{f10} \\
f_{\psi} &= A \pr + B \pr c_{\alpha}, \, \, \alpha=78 \label{fpsi}
\end{align}

\noindent and for the SO(10) breaking, we have 

\begin{align}
\alpha_{10} &= \alpha_{5} f_{5} = \alpha_{\chi} f_{\chi} \label{SO(10)couplings} \\
f_{5} &= A^{\prime \prime} + B^{\prime \prime} c_{\alpha}, \, \, \alpha=(1,...,24) \label{f5} \\
f_{\psi} &= A^{\prime \prime} + B^{\prime \prime} c_{\alpha}, \, \, \alpha=45 \label{fchi}
\end{align}

\noindent For both $E_6$ and SO(10) (and indeed for any simple Lie algebra with the exception of SU(N), $N\geq3$) the $d_{\alpha \beta \gamma}$ and hence the $c_{\alpha}$ are zero \cite{Georgi:1999la}.  Accordingly, we can take the $B \pr, B^{\prime \prime}$ in Eqs. (\ref{f10}, \ref{fpsi}, \ref{f5}, \ref{fchi}) to be zero.  Matching with Eq. (\ref{xyAB}) of the SU(5) case, this clearly leads to $x=0$, and Eq. (\ref{gcMU}) shows that this corresponds to no relative splitting of the gauge couplings at unification.  We can, however, have a shift by the parameter y in all the couplings after each breaking.  These parameters will depend on the flux breaking mechanism, and we will leave them as free parameters of the model:

\begin{align}
\frac{1}{\alpha_{10}} &=  \frac{1}{\alpha_{6}} - y \pr \notag \\
\frac{1}{\alpha_{\psi}} &=  \frac{1}{\alpha_{6}} - y \pr \notag \\
\frac{1}{\alpha_{5}} &=  \frac{1}{\alpha_{6}} - y^{\prime \prime}  \notag \\
\frac{1}{\alpha_{\chi}} &=  \frac{1}{\alpha_{6}} - y^{\prime \prime}  \label{yprime}
\end{align}

\noindent With $\alpha_G = \alpha_5$ in Eq. (\ref{gcMU}), we can then proceed with the analysis as for the SU(5) case.  It should be noted that in Eq. (\ref{yprime}), the signs of y and $y \pr$ are not known, and so the $U(1)_N$ gauge coupling could be either bigger or smaller than $\alpha_5$ at unification.  This splitting is a free parameter of the model.  

\subsection{The Spectrum, and One Loop Renormalisation Group Analysis}

In the considered model we have the following vector pairs of exotics, which get large masses when $\theta_{31}$ gets a VEV: ($d+ \ov{d}^c$), ($Q+ \ov{Q}$), ($H_d + \ov{H}_d$), 2($L+ \ov{L}$), 2($u^c + \ov{u}^c$).  Below some scale $M_X < M_{GUT}$ these exotics decouple.  We then have the extra exotics 3($D+ \ov{D}$), 2($H_u , H_d$) which survive to low energy and decouple at a scale $M_{X \pr}=1TeV$.  Below the scale $M_{X \pr}$, we have the low energy matter content of the MSSM.
The low energy values of the  gauge couplings are given by the evolution equations
\be\label{Brun}
\frac{1}{\alpha_a(M_Z)} = \frac{1}{\alpha_{a}(M_{GUT})}+\frac{b_a^x}{2\pi}\,\ln\frac{M_{GUT}}{M_X}+ \frac{b_a^{x\pr}}{2\pi}\,\ln\frac{M_{X}}{M_{X \pr}}+\frac{b_a}{2\pi}\,\ln\frac{M_{X \pr}}{M_Z}
\ee
where $b_a^x$ is the beta-function above the scale $M_X$,  $b_a^{x \pr}$ is the beta-function below $M_X$ and $b_a$ is the beta-function below $M_{X \pr}$.  Combining  the above equations, we find that the GUT scale is given by
\be
M_{GUT} =  e^{\frac{2\pi}{\beta {\cal A}}\rho}\, M_Z^{\rho} M_{X \pr}^{\gamma - \rho} M_X^{1-\gamma}
\label{M_U}
\ee
where ${\cal A}$ is a function of the experimentally known low energy values of the
SM gauge coupling constants
\ba
\frac{1}{\cal A} &=& \frac 53 \,\frac{1}{\alpha_1(M_Z)}-\frac{1}{\alpha_2(M_Z)}-\frac 23 \frac{1}{\alpha_3(M_Z)}
\nn\\
&=&\frac{\cos(2\theta_W)}{\alpha_{em}}-\frac 23 \frac{1}{\alpha_3(M_Z)}
\ea
We have also introduced  the ratios $\rho$ and $\gamma$
\be
\rho  = \frac{\beta}{\beta_x} \, \, \gamma = \frac{\beta_{x \pr}}{\beta_x}
\ee
where $\beta,\beta_{x \pr} , \beta_{x}$ are the beta-function combinations in the regions $M_Z < \mu < M_{X \pr}$, $M_{X \pr} < \mu < M_{X}$ and $M_X < \mu < M_{GUT}$ respectively
\begin{align}
\beta_x&=b_Y^x-b_2^x-\frac 23b_3^x \label{betax2}\\
\beta_{x \pr}&=b_Y^{x \pr}-b_2^{x \pr}-\frac 23b_3^{x \pr} \label{betaxpr2}\\
\beta&=b_Y-b_2-\frac 23b_3 \label{beta02}
\end{align}
\noindent The beta function coefficients are given by ($b_1=\frac 35\, b_Y$)
\ba
b_1&=&-0+2 n_f+\frac{3}{10}(n_h+n_L)+\frac{1}{5}n_{d^c}+\frac{1}{10}n_Q+\frac{4}{5}n_{u^c}
+\frac 35\,n_{e^c} \label{b1}
\\
b_2&=&-6+2n_f+\frac 12 (n_h+n_L)+0\,n_{d^c}+\frac 32\,n_Q+0\,n_{u^c} \label{b2}
\\
b_3&=&-9+2 n_f+0\,(n_h+n_L)+\frac 12\,n_{d^c}+n_Q+\frac 12\,n_{u^c} \label{b3}
\ea
with $n_f=3$ the number of families and $n_{h,L,...}$ counting  Higgses and exotic matter.  For our spectrum, the coefficients are given by

\begin{align}
b_1 &= 6.6, \, \, b_2 = 1, \, \, b_3 = -3 \label{bac} \\
b_1^{x \pr} &= 9, \, \, b_2 = 3, \, \, b_3 = 0 \label{baxpc} \\
b_1^x &= 14.6, \, \, b_2 = 9, \, \, b_3 = 5 \label{baxc}
\end{align}

\noindent Plugging these numbers into Eq. (\ref{M_U}), we see  that $M_{GUT}$ becomes independent of
the $M_X$ and $M_{X'}$ scales and in fact it is identified with the MSSM unification scale
\begin{equation}
M_U=M_{GUT}\equiv e^{\frac{2\pi}{\beta {\cal A}}}\,M_Z\approx 2\times 10^{16}\textrm{GeV} \label{mgut}
\end{equation}

\subsection{Model Dependence of the Splitting Parameter, x}

From Eq. (\ref{gcMU}), the splitting of the standard model gauge couplings is given by

\begin{equation}
x=\frac{1}{\alpha_2 (M_G )}- \frac{1}{\alpha_3 (M_G )}
\end{equation}

\noindent We can now use the evolution equation (\ref{Brun}) to relate x to the low energy coupling constants $\alpha_2$ and $\alpha_3$, giving

\begin{equation}
\left(\frac{1}{\alpha_2}-\frac{1}{\alpha_3}\right)_{M_Z}= x+ \frac{b_2^x-b_3^x}{2\pi}\log\left(\frac{M_G}{M_X}\right)
+ \frac{b_2^{x \pr}-b_3^{x \pr}}{2\pi}\log\left(\frac{M_X}{M_{X \pr}}\right) \frac{b_2-b_3}{2\pi}\log\left(\frac{M_{X \pr}}{M_Z}\right) 
\end{equation}

\noindent Using Eqs. (\ref{bac}, \ref{baxpc}, \ref{baxc}, \ref{mgut}) and the relations $\alpha_{em}=\alpha_{2} sin^2 \theta_{w}$, $\frac{1}{\alpha_Y}=\frac{(1-sin^2 \theta_{w})}{\alpha_{em}}$ and $\alpha_1 = \frac{5}{3} \alpha_Y$, we arrive at the following expression for x

\begin{align}
x &=\frac 43\frac{1}{\alpha_2}-\frac 13\frac{1}{\alpha_Y}-\frac 79\frac{1}{\alpha_3}-\frac{1}{2 \pi} \ln\left({\frac{M_{x \pr}}{M_x}}\right) \notag \\
&= \frac{(5 sin^2 \theta_{w} - 1)}{3 \alpha_{em}} - \frac{7}{9} \frac{1}{\alpha_3} - \frac{1}{2 \pi} \ln\left({\frac{M_{x \pr}}{M_x}}\right) \label{xeqn}
\end{align}

\noindent It can be seen that the factors which affect the splitting are the matter content of the spectrum (which manifests itself in the numbers multiplying the Standard Model parameters), and the ratio of the two exotic mass scales.  At this point, we can compare the E6SSM model with the E6 based model of \cite{Callaghan:2011jj} (model 1), where the E6SSM light exotics are heavy.  We can use the above equation for both models as they have the same spectrum, the difference being in the scales at which the exotics decouple.  In the E6SSM case we have $M_{X \pr} = 1TeV$ and from the calculated singlet VEVs, $M_X = 1.44 \times 10^{16} GeV$, whilst in model 1, we have $M_{X \pr} = 0.306 \times 10^{16} GeV$ and $M_X = 1.31 \times 10^{16} GeV$.  Taking the values $\alpha_{em}^{-1}(M_Z ) = 127.916$, $sin^2 \theta_{w} (M_Z ) = 0.23116$ and $\alpha_3 = 0.1184$, the part of the right hand side of Eq. (\ref{xeqn}) involving these parameters is evaluated as 0.07.  Due to the fact that this number is small, in order for x to be close to zero (corresponding to the usual case of gauge coupling unification) the masses of both sets of exotics need to be close together.  This is the case in model 1 where we have x=0.3, but not in the case of the E6SSM model where x=4.9.  

\begin{figure}
    \includegraphics[height=3in, width=4in]{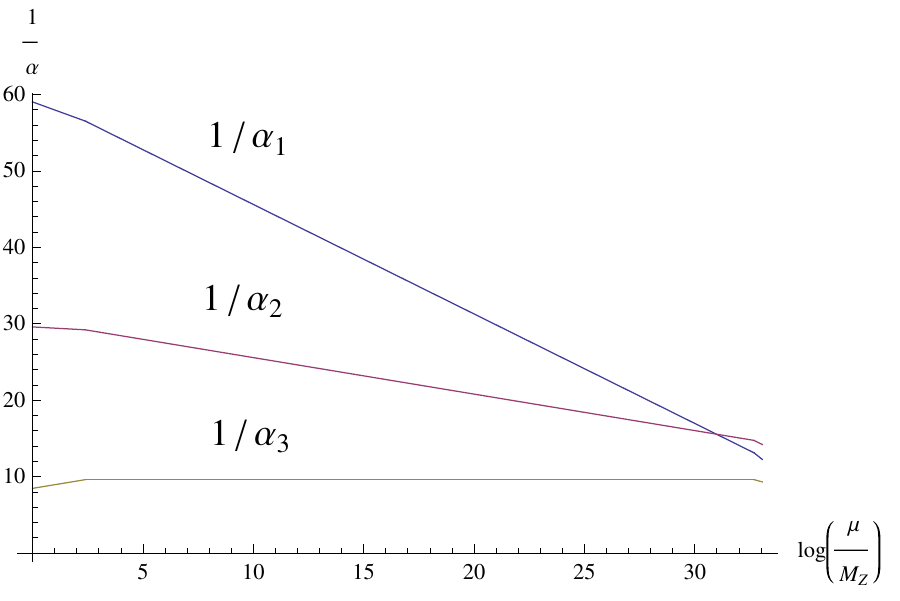} 
  \caption{The running of $\alpha_1$, $\alpha_2$ and $\alpha_3$ from their SM value at $M_Z$ up to $M_{GUT}$ for the case of the F-theory E6SSM} \label{RGE E6SSM}
\end{figure}

\begin{figure}
    \includegraphics[height=3in, width=4in]{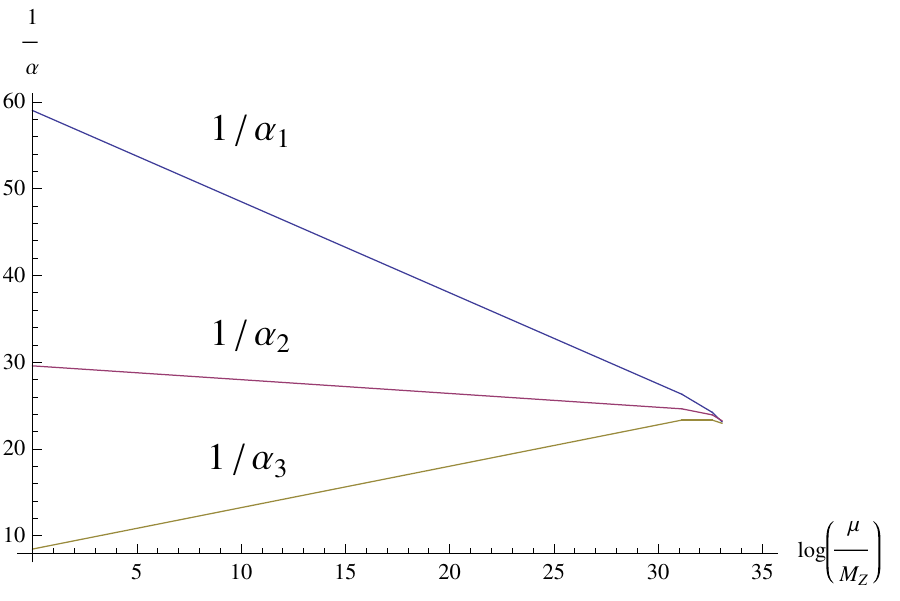} 
  \caption{The running of $\alpha_1$, $\alpha_2$ and $\alpha_3$ from their SM value at $M_Z$ up to $M_{GUT}$ for the case of model 1, presented in \cite{Callaghan:2011jj}} \label{RGE model 1}
\end{figure}

Taking the low energy values of $\alpha_1$, $\alpha_2$ and $\alpha_3$ and using the one loop remormalisation group equations to run the couplings up to the unification scale (taking into account the presence of the exotic matter) results in Figure \ref{RGE E6SSM} for the F-theory E6SSM, and Figure \ref{RGE model 1} for model 1.  In Figure \ref{RGE E6SSM}, the reciprocals of the gauge couplings are split by approximately 35 percent (relative to the largest value) at unification, whereas in Figure \ref{RGE model 1} they meet to 1.3 percent accuracy.  The fact that the gauge couplings meet in model 1 means that our spectrum is special for the case of heavy exotics.  If we want the couplings to unify in the F-theory E6SSM, it may be possible to change the spectrum, adding in extra exotics which modify the renormalisation group running.

In \cite{Callaghan:2011jj}, we used the fact that $x>0$ in order to obtain a lower bound on $\alpha_3$ (given the low energy experimental values of $\alpha_1$ and $\alpha_2$ as input parameters).  As x is close to zero in this model, the result is near the limiting case, and the bound is $\alpha_3 \geq 0.113$, consistent with (but quite close to) the experimental value.  Repeating the calculation for the E6SSM gives

\begin{equation}
\alpha_3 \ge \frac{7}{9}\frac{1}{\frac{5 \,{\sin^2\theta_W}-1}{3 \,\alpha_e}
-\frac{1}{2 \pi} \ln\left({\frac{M_{x \pr}}{M_x}}\right)} \approx 0.068
\end{equation}

\noindent As such, we have a bound which is consistent with experiment, and much less stringent than that of model 1.

\subsection{Baryon- and lepton-number violating terms}
As discussed above the R-parity violating superpotential couplings $u^cd^cd^c$, $Qd^cL$, $Le^cL$, $\kappa LH_u$ are not allowed because of the underlying $U(1)$ symmetries which play the role of R-parity.
Dimension 5 terms in the Lagrangian,
corresponding to the superpotential terms $QQQL$ and $u^{c}u^{c}d^{c}e^{c}$,
which would be allowed by usual R-parity, are forbidden by the $U(1)$ symmetries that originate in the underlying $E_{6}$.

\begin{figure}
\centering
\includegraphics[scale=.3,angle=0]{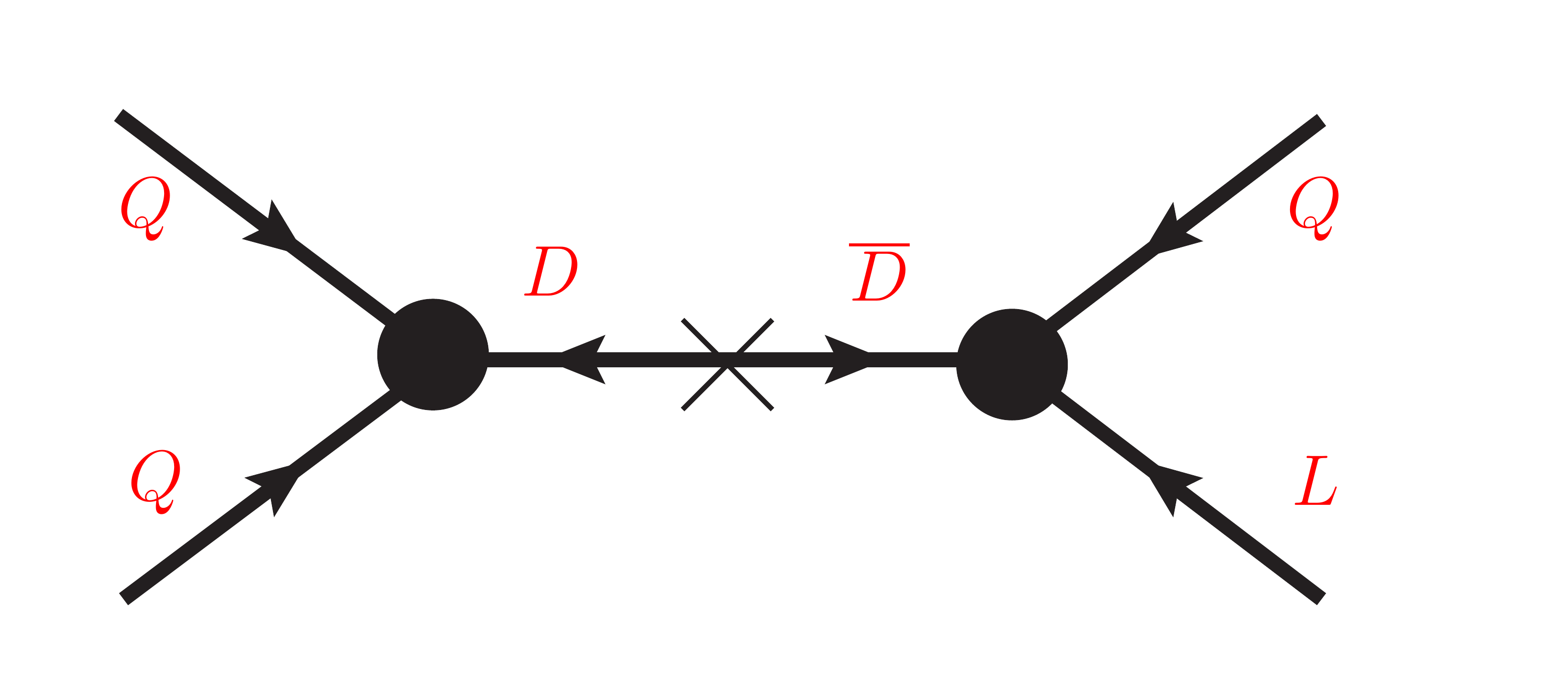}
\caption{\small{The general proton decay diagram generating the dimension 5 operator $QQQL$.}
} \label{PD0}
\end{figure}

\begin{figure}[!b]
\centering
\includegraphics[scale=.7,angle=0]{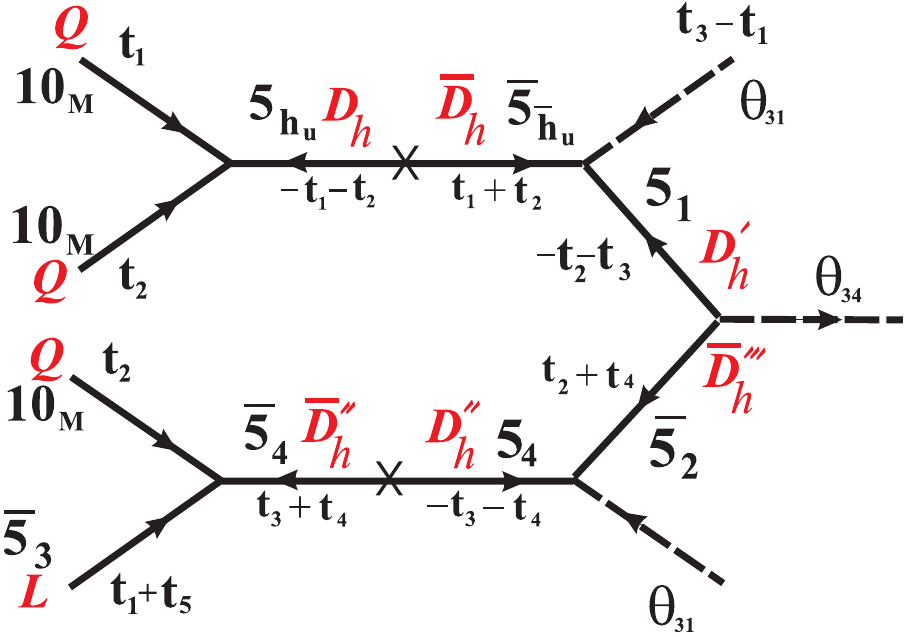}
\caption{\small{The specific proton decay diagram generating the dimension 5 operator $QQQL$ in this model.}
} \label{PD}
\end{figure}

Of course one must be careful that spontaneous symmetry breaking terms coming from SM singlet field VEVs do not allow these dangerous operators to appear.  Allowing for arbitrary singlet fields to acquire VEVs the dangerous the baryon- and lepton-number violating operators arise through the terms  $\theta_{15}LH_{u}$, $(\theta_{31}\theta_{45}+\theta_{41}\theta_{35})10_M\overline{5_3}^2$ and $\theta_{31}\theta_{41}10_M^3\overline{5_3}$. Thus, provided $\theta_{15}$, $\theta_{41}$ and $\theta_{45}$ do {\it not} acquire VEVs these dangerous terms will not arise.

However this is not sufficient to ensure the absence of baryon and lepton number violating terms because, even in the absence of these VEVs, tree level graphs can generate the dangerous operators at higher order in the singlet fields.  As such, we must look for graphs of the type shown in Fig.~\ref{PD0}.  In these models, the dangerous graph is shown in Fig.~\ref{PD} and is driven by colour triplet exchange coming from the couplings
\ba
10_M\,10_M\,5_{H_u}&\ra & QQD_h +\ldots  \nn\\
 5_{H_u}\bar 5_{\bar H_u}&\ra& M_DD_h\bar D_h +\ldots  \nn
 \\
 \theta_{34}5_1\bar 5_2&\ra&  \langle\theta_{34}\rangle\,D_h'\bar D_h{'''} +\ldots   =
 \langle\theta_{34}\rangle\,D\bar D +\ldots        \nn .
\ea  
The notation has been simplified here by calling the light exotics $D_h'$ and $\bar D_h{'''}$ simply $D$ and $\bar{D}$.  In Fig.~\ref{PD} the full notation is used, but in Fig.~\ref{PD0} and Fig.~\ref{forbidden} the simplified notation is used, with D representing a light colour triplet. 

As may be seen from Table~\ref{1} 
only the states $D$ and $\bar{D}$
(i.e. $D'_{h}$ and $\bar{D}_{h}'''$ in Fig.~\ref{PD})
appear in the spectrum with mass generated by the singlet VEV $ \langle\theta_{34}\rangle$ which is at the TeV scale. 
Since the choice of fluxes in Table~\ref{1} eliminates
light colour triplet states $D_h$ in the low energy spectrum, arising from $5_{H_u}$, there is
no reason to expect any KK modes with the quantum numbers of $D_h$ below the string scale
since there is no ground state with the colour triplet quantum numbers of $D_h$ below the string scale.
Similarly the choice of fluxes in Table~\ref{1} eliminates
light colour triplet states $D_h{''}$ in the low energy spectrum, arising from $5_{4}$, so there is
no reason to expect any KK modes with the quantum numbers of $D_h{''}$ below the string scale.

If string states with the quantum numbers of $D_{h},D_{h}''$ exist they are expected to have string scale masses, of $O(M_{S})$. In this case the diagram of Fig.~\ref{PD} gives the proton decay operator $QQQL$ with coefficient  $1/\Lambda_{eff}$ given by
\ba\frac{1}{\Lambda_{eff}}=\lambda^5\,\left(\frac{\langle\theta_{31}\rangle}{M_S}\right)^2
\frac{1}{\langle\theta_{34}\rangle}\label{lambda}
\ea
In (\ref{lambda}), $\lambda^5$ represents the the product of the five
Yukawa couplings in the relevant diagram and according to ref~\cite{Leontaris:2010zd}
it is expected to be
\[\lambda^5=\lambda_{10\cdot 10\cdot 5}\lambda_{10\cdot \bar 5\cdot\bar 5}\lambda_{5\cdot\bar 5\cdot 1}^3\approx 10^{-3}.\]

This implies
\[\Lambda_{eff} \approx 10^3\,\left(\frac{M_S}{\langle\theta_{31}\rangle}\right)^2
\langle\theta_{34}\rangle \, .\]

This, multiplied by the appropriate loop-factor due to higgsino/gaugino dressing
and other theoretical factors~\cite{Murayama:1994tc,Babu:1998wi,Goto:1998qg,Dermisek:2000hr,Murayama:2001ur},
 should be compared to experimental bounds on nucleon decay. This bound, relevant to the case that the operator $QQQL$ involves quarks from the two  lighter generations only,  requires $\Lambda_{eff}^{light}>(10^{8}-10^{9})M_{S}$.
 Since $\langle\theta_{34}\rangle \sim TeV \ll M_S$, there will be a large discrepancy between  $\Lambda_{eff}^{light}$ and $\Lambda_{eff}$, even when the suppression factors for the first and second generations (due to non perturbative flux corrections) are considered \cite{Callaghan:2011jj}.  As such, it is clearly
 necessary to forbid the light quark operator generated by the diagram of Figure \ref{PD}.
  One way to do this would be to forbid the coupling $\theta_{31}\overline{5}_{\overline{H}_U}5_1$. 
  Note that all the other vertices in Figure \ref{PD} are necessary for various phenomenological reasons.
  For example, the couplings in Figure \ref{botm} are necessary to generate the bottom quark Yukawa coupling,
  and so these couplings cannot be set to zero. Similarly the top quark Yukawa coupling originates from the coupling
  $10_M10_M5_{H_U}$. The coupling $\theta_{34} 5_1\overline{5}_2$ is necessary to give the exotics
  a TeV scale mass term $\langle\theta_{34}\rangle D\overline{D}$ .
  
In fact we only need to 
forbid the colour triplet components of the $\theta_{31}\overline{5}_{\overline{H}_U}5_1$ coupling.
This can be achieved by 
imposing a discrete $Z_2$ symmetry with 
the following set of fields chosen to be odd: 
($D_{h} \pr$, 
$\ov{D}_{h}^{\prime \prime \prime}$, 
$D_{h}^{\prime \prime}$, $\ov{D}_{h}^{\prime \prime} $). 
Either the set ($L,e^c$) or ($Q,d^c,u^c$)
are also chosen to be odd. All other fields are chosen to be even under  $Z_2$.
These assignments forbid the proton decay diagram in Fig.~\ref{PD} but allow the top quark Yukawa coupling.


Note that with these charge assignments the $Z_2$ symmetry is absolutely conserved.
Also $Z_2$ doesn't respect SU(5), as for example $\ov{D}_{h}^{\prime \prime} (\ov{5}_{4})$ must be odd, but the $\ov{H}_{d}$ state coming from the same curve must be even.  This is because it gets a large mass from the coupling $\theta_{31} H_{d} \ov{H}_{d}$, and the $\theta_{31}$ and $H_{d}$ fields must be even otherwise $Z_2$ would be broken leading to cosmological domain walls. 
The $Z_2$ symmetry clearly goes beyond the rules of local F-theory, which corresponds to the fact that we are appealing to global F-theory to forbid 
the colour triplet components of the $\theta_{31}\overline{5}_{\overline{H}_U}5_1$ coupling by a 
geometric suppression mechanism.  However, in the present paper this just corresponds to an 
assumption related to the uncertain nature of singlet fields and their couplings in F-theory.
Such assumptions about singlets are always required in any case.  In particular, the forbidden coupling involves $\theta_{31}$ which doesn't live in a 27 of E6, and the Yukawa couplings of such singlets are particularly poorly understood. \footnote{Note that the $\theta_{14}$ and $\theta_{34}$ are different types of singlet since they are contained in 27s of E6 and hence have matter curves.}

\section{Comparison with known models}

\subsection{E6SSM}

The low energy spectrum in Table~\ref{1} resembles that of the standard E6SSM \cite{King:2005jy,King:2005my,King:2007uj}. The F-theory model with a surviving Abelian gauge group is also a supersymmetric standard model 
involving the same $U(1)_N$ gauge symmetry surviving down to the TeV scale. However, whereas the E6SSM matter content appears to arise from three 27 representations of $E_6$, in the F-theory model there is a rather subtle
doublet-triplet splitting involved in achieving this spectrum, due to the effects of flux,
as indicated in Table~\ref{1}. The light exotics with the quantum numbers of colour triplets $D$ and $\overline{D}$
arise from three $5_1$ and three $\ov{5}_2$ representations of $SU(5)$, while the third Higgs doublet
$H_u$ arises from a different representation $5_{H_u}$. 

\begin{figure}[!b]
\centering
\includegraphics[scale=.2,angle=0]{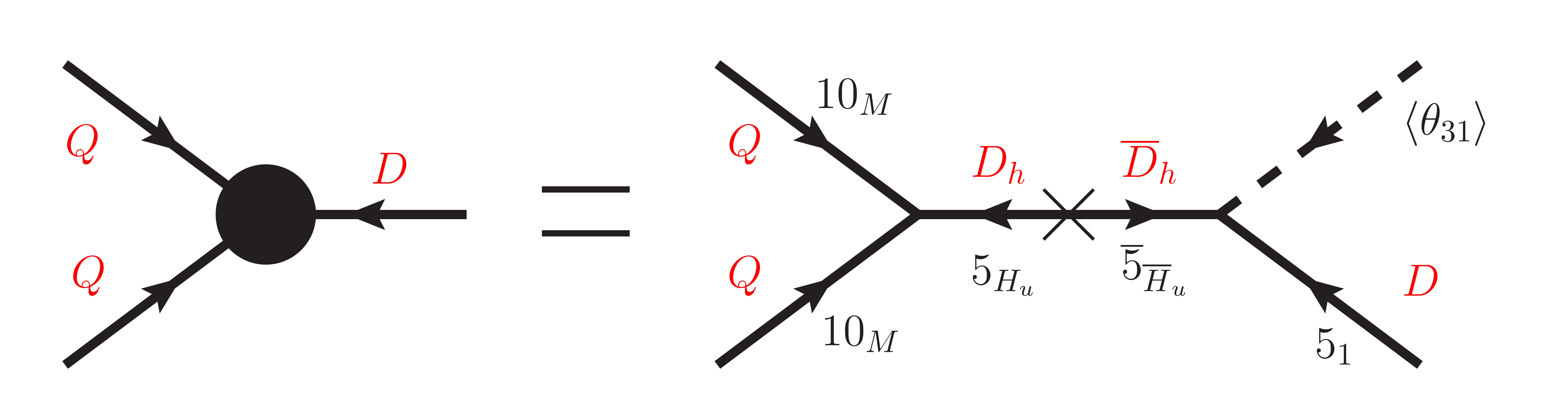}
\caption{\small{Coupling $DQQ$ forbidden by the imposed $Z_2$ symmetry, where the field $D$ is a TeV scale exotic.}
} \label{forbidden}
\end{figure}

\begin{figure}[!b]
\centering
\includegraphics[scale=.2,angle=0]{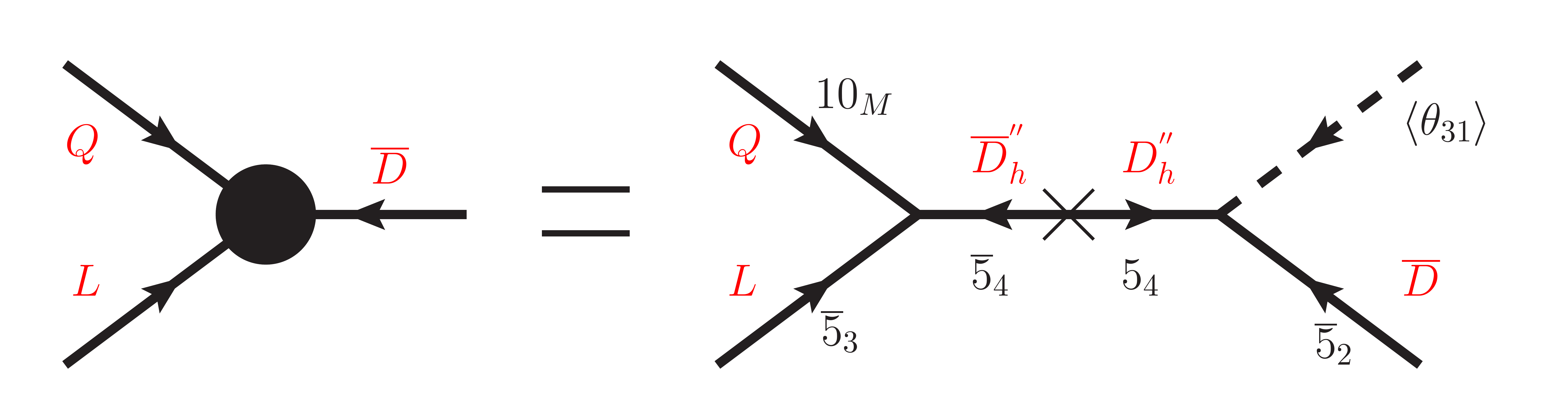}
\caption{\small{Coupling $\ov{D}QL$ allowed by the imposed $Z_2$ symmetry, where the field $\ov{D}$ is a TeV scale exotic.}
} \label{allowed}
\end{figure}

The low energy gauge invariant superpotential of the E6SSM can be written
\be
W^{\rr{E}_6\rr{SSM}} = W_0 + W_{1,2},\label{eq:w}
\ee
where $W_{0,1,2}$ are given by
\bea
W_0 &=& W_{\rr{Yukawa}} +
\lambda_{ijk}\hat{S}_i\hat{H}_{dj}\hat{H}_{uk} + \kappa_{ijk}\hat{S}_i\hat{\bar{D}}_j\hat{D}_k,
\label{W0}\\
W_1 &=& g^Q_{ijk}\hat{D}_i\hat{Q}_{Lj}\hat{Q}_{Lk} + g^q_{ijk}\hat{\bar{D}}_i\hat{d}_{Rj}^c\hat{u}_{Rk}^c,\label{eq:w1}\\
W_2 &=& g^N_{ijk}\hat{N}^c_i\hat{D}_j\hat{d}_{Rk}^c + g^E_{ijk}\hat{D}_i\hat{u}_{Rj}^c\hat{e}_{Rk}^c + g^D_{ijk}\hat{\bar{D}}_i\hat{Q}_{Lj}\hat{L}_{Lk}.\label{eq:w2}
\eea
with $W_{1,2}$ referring to either $W_1$ or $W_2$, giving two alternative models in the usual E6SSM.
In the E6SSM the three $SU(5)$ singlets $S_i$ which are charged under $U(1)_N$ may be labelled as 
$S_{\alpha}, \, \alpha=1,2$ and $S_3$, where the latter couples to exotics, giving them mass and generating the 
effective $\mu$ term after they acquires a non zero VEV.  
In the F-theory model these are identified as two copies of $\theta_{34}$ which give the light exotics mass, and the $\theta_{14}$ which generates the $\mu$ term in the F-theory model. The other GUT singlets which get VEVs in the F-theory model are $\theta_{31}$ (which removes unwanted exotics from the low energy spectrum), and $\theta_{53}$ (which helps generate neutrino masses).  These singlets acquire string scale VEVs, and are uncharged under the $U(1)_N$ as required.  The other important singlet is $\theta_{12}$, as this is the Majorana state which we call $S_R$.  This singlet is uncharged under the perpendicular U(1)s and so can get a Majorana mass and play a role in the double see-saw mechanism for generating neutrino masses.  

Another difference between the models is that in the E6SSM there are the $H \pr , \, \ov{H} \pr$ states coming from an incomplete 27 and $\ov{27}$ representation, which are necessary to ensure gauge coupling unification.  In F-theory however, we have splitting of the gauge couplings at unification as discussed, and so these extra fields are not needed.  As such, the model presented in this paper more closely resembles the ME6SSM of \cite{Howl:2007zi}.  
However, in the F-theory model, no intermediate Pati-Salam gauge group is required.
Due to the splitting of the couplings at unification, we cannot know about the size of the $U(1)_N$ gauge coupling.  As the normal limits on the $Z \pr$ come from the assumption of unification, these limits do not apply in the F-theory model.   

It should be noted that in the local F-theory version of the E6SSM all the couplings of Eqs.\ref{eq:w1} and \ref{eq:w2} are forbidden at the level of renormalisable operators due to the perpendicular U(1)s.
At the level of local F-theory, they are all allowed at the effective level after including one insertion of the 
$\theta_{31}$ field. However at the level of global F-theory we have assumed that not all couplings 
involving $\theta_{31}$ are allowed, and we have described this by imposing a $Z_2$ symmetry
so that certain effective diagrams involving the exchange of heavy colour triplet states
are forbidden, in particular those which would lead to proton decay. 

The effective $DQQ$ coupling is forbidden by $Z_2$ since $D$ is odd.
In detail, the reason why this operator is forbidden is shown in Fig.~\ref{forbidden}
since $D$ is odd and ($D_{h} $, 
$\ov{D}_{h}$) are both even. Note that the $Z_2$ symmetry that we imposed
has a global F-theory interpretation as being due to a geometrically suppressed $\theta_{31}$ vertex.  
Similar arguments would forbid the effective $D u^c e^c$ coupling being generated by a diagram
analogous to Fig.~\ref{forbidden}.
Note that even though a renormalisable $D u^c e^c$ operator would be allowed by $Z_2$,
it is forbidden by the rules of local F-theory.

On the other hand the $\ov{D}QL$ coupling is allowed by $Z_2$ and 
can be generated effectively by non-renormalisable operators as shown in Fig.~\ref{allowed}.
All couplings in this diagram are allowed by $Z_2$
since $\ov{D}$ is odd and in this case also 
($D_{h}^{\prime \prime}$, $\ov{D}_{h}^{\prime \prime} $) are odd,
as is the combination $QL$.
Thus the effective coupling $\ov{D}QL$ is successfully generated, allowing the  $\ov{D}$ to decay as a chiral leptoquark with couplings to left-handed quarks and leptons. Note that the effective $\ov{D}d^cu^c$ coupling is forbidden by $Z_2$ since 
$\ov{D}$ is odd while the combination $d^cu^c$ is even.

By contrast in the ME6SSM all the couplings involving $D$ and $\overline{D}$ are all highly suppressed coefficients.
This tends to give long lived $D$ decays in the ME6SSM, but prompt $D$ decays in the F-theory model,
with large couplings to left handed quarks and leptons, providing characteristic and striking signatures at the LHC.

In summary, proton decay is suppressed by the geometric coupling suppression of a singlet state $\theta_{31}$, 
which we interpret in terms of a $Z_2$ symmetry. This symmetry 
effectively forbids all the couplings of the exotic charge $-1/3$ colour triplet state $D$ to quarks and leptons, while allowing the coupling involving $\overline{D}QL$. However the coupling $\overline{D}d^cu^c$ is forbidden by $Z_2$.
Thus $\overline{D}$ decays as a chiral leptoquark with couplings to left-handed quarks and leptons, with $D$ coupling to $\overline{D}$ to make a TeV scale Dirac fermion.
We emphasise again that the effective coupling 
$Du^ce^c$ is forbidden, while $\overline{D}QL$ is allowed 
providing a distinctive signature of chiral leptoquarks.

\subsection{NMSSM+}

The low energy spectrum in Table~\ref{1} may also apply to a version of the F-theory model in which there is no
additional Abelian gauge group present, in other words where the $U(1)_N$ gauge group is broken by flux at the GUT scale. This was the case for the F-theory model in \cite{Callaghan:2011jj}.
The difference between the present F-theory model and that in \cite{Callaghan:2011jj}
is then mainly in the order of magnitude of the  
the singlet $\theta_{34}$ VEV as determined by the different flatness conditions in the two models.
In the previous model the singlet $\theta_{34}$ acquired a string scale VEV which gave large masses to the exotic states.
In the present model the singlet $\theta_{34}$ 
acquires a TeV scale VEV which remain light in the current model. It was also assumed in \cite{Callaghan:2011jj} that the $\mu$ term is generated when the U(1) symmetries are explicitly broken by non-perturbative effects.
Here we assume that the singlet $\theta_{14}$ acquires an electroweak scale VEV which generates an effective
$\mu$ term. There will also be non-perturbative corrections which generate trilinear
self-couplings and additional electroweak scale masses for 
$\theta_{14}$, explicitly breaking all global U(1) symmetries.

The resulting F-theory model with the spectrum in Table~\ref{1} but with no additional Abelian gauge group present,
resembles that of the NMSSM+ \cite{Hall:2012mx}. However in the F-theory model 
the $U(1)_N$ is broken by flux at a high scale, whereas in the NMSSM+ it is broken by an explicit sector.
Recall that the usual NMSSM is based on the scale invariant superpotential \cite{Ellwanger:2009dp},
\begin{equation}
W_{\rr{NMSSM}} = W_{\rr{Yukawa}} + \lambda S H_u H_d + \frac{1}{3} \kappa S^3,
\end{equation}
where $W_{\rr{Yukawa}}$ represents the MSSM Yukawa couplings.
In the F-theory model we identify the singlet $S$ of the NMSSM
with $\theta_{14}$. The trilinear self-coupling and other linear and quadratic terms are generated by 
non-perturbative corrections, resulting in a 
generalised NMSSM (GNMSSM) \cite{King:1995ys, Ross:2012nr} with superpotential,
\begin{equation}
W_{GNMSSM} = W_{\rr{Yukawa}} + (\mu +\lambda S) H_u H_d + \frac{1}{2} \mu_{s} S^2 + \frac{1}{3} \kappa S^3, \label{GNMSSM}
\end{equation}
where the singlet $S$ of the GNMSSM is again identified
with $\theta_{14}$.
The non-perturbative corrections responsible for these terms are similar to those 
which were used to generate the $\mu$ term in \cite{Callaghan:2011jj}.  

However the model is more than the usual GNMSSM since it also involves
the exotic sector of the NMSSM+, so it more closely resembles a sort of GNMSSM+
with three compete 27 dimensional families
\cite{Hall:2012mx}. The superpotential terms involving the other exotic states
(apart from  $\theta_{14}$) are similar to those of the E6SSM in Eq.\ref{eq:w}
and discussed in the preceding subsection.
The phenomenological comments also discussed in the preceding subsection
concerning unification, proton decay and the $\overline{D}$ couplings at the LHC all apply to this case as well
where the $U(1)_N$ is broken. The main advantage of the NMSSM+ over the E6SSM is that the fine-tuning is lower due to the absence of $U(1)_N$ D-terms which would introduce a term in the Higgs potential proportional to the fourth power of the $Z'$ mass as discussed in \cite{Hall:2012mx}. The NMSSM+, involving  three compete 27 dimensional families, has lower fine-tuning than the NMSSM, which in turn has lower fine-tuning than the MSSM \cite{Hall:2012mx}, making it the lowest fine-tuned model consistent with perturbative unification.

\section{Summary and Discussion}
In this paper we have explored F-theory models in which the low energy supersymmetric theory
contains the particle content of three 27 dimensional representations of the underlying $E_6$ gauge group,
plus two extra right-handed neutrinos predicted from F and D flatness.
Using the techniques of semi-local model building in F-theory,
we have shown that it is possible to formulate F-theory models whose 
TeV scale effective theory resembles either the E6SSM or the 
NMSSM+, depending on whether an additional Abelian gauge group does or does not survive.
However there are novel features compared to both these models as follows:
\begin{enumerate}
\item If the additional Abelian gauge group is unbroken then it can have a weaker gauge
coupling than in the E6SSM.
\item If the additional Abelian gauge group is broken then non-perturbative effects can violate
the scale invariance of the NMSSM+ leading to a generalised model.
\item Unification is achieved not at the field theory level but at the F-theory level 
since the gauge couplings are split by flux effects, negating the need for 
any additional doublet states which are usually required.
\item Proton decay is suppressed by the geometric coupling suppression of a singlet state, 
which is possible in F-theory, which effectively suppresses the coupling of the exotic charge $-1/3$ colour triplet state $D$ to quarks and leptons.
\item The $\overline{D}$ decays as a chiral leptoquark with couplings to left-handed quarks and leptons, providing characteristic and striking signatures at the LHC.
\end{enumerate}

\begin{table}[htdp]
\small
\centering
\begin{tabular}{|c|c|c|c|}
\hline
Model Features& F-MSSM & F-E6SSM  & F-NMSSM+ \\
\hline
$\left\langle \theta_{53} \right\rangle$, $\left\langle \theta_{31} \right\rangle$& $\sim M_X$ & $\sim M_X$  & $\sim M_X$ \\
\hline
$\left\langle \theta_{34} \right\rangle$ & $\sim M_X$ & $\sim$ 1 TeV  & $\sim$ 1 TeV \\
\hline
$\left\langle \theta_{14} \right\rangle$ & 0 &  $\sim$ 1 TeV  & $\sim$ 1 TeV  \\
\hline
$U(1)_N$ breaking & Flux $\sim M_X$ & $\left\langle \theta_{34} \right\rangle \sim$ 1TeV  & Flux $\sim M_X$\\
\hline
Non perturbative $\mu$ term & $\mu^{N.P} H_u H_d$ & -  & - \\
\hline
Effective $\mu$ term & - &  $\theta_{14} H_u H_d$ & $\theta_{14} H_u H_d$ \\
\hline
Non perturbative singlet masses& - & -  & $m_s \theta_{14}^2$, $m_s^2 \theta_{14}$ \\
\hline
\end{tabular}
\caption{\small Similarities and differences between different F-theory based models which go beyond the MSSM.}
\label{models table}
\end{table}%

The particle spectrum of the F-theory models is summarized in Table~\ref{1}.
The models here may be compared to the F-theory model in \cite{Callaghan:2011jj}
in which the singlets $\theta_{34}$ acquired a string scale VEV which gave large masses to the exotic states,
yielding a low energy theory as in the MSSM, which we can call an F-MSSM. 
The new models here have a singlet spectrum where the new flatness conditions allow 
the singlets $\theta_{34}$ to have small VEVs resulting in a light exotic mass spectrum.
In addition the singlets $\theta_{14}$ are used to generate electroweak scale effective
$\mu$ terms.  Five right handed neutrinos, as well as other restrictions on the numbers of certain singlets in the spectrum, are required to make the model consistent with F and D-flatness conditions.  
If the gauged $U(1)_N$ is broken by flux at the GUT scale then we have either the F-MSSM
as discussed previously or the F-NMSSM+ as investigated here, where non-perturbative corrections
break all global $U(1)$ symmetries via $\theta_{14}$ mass terms. However if the gauged $U(1)_N$
is unbroken then we are led to an F-E6SSM but with the phenomenological differences discussed above.
For example, we emphasise that unification in the F-MSSM \cite{Callaghan:2011jj} 
may be achieved approximately at the field theory level,
since the exotic states occur at high energy and have a small mass splitting, 
while in the F-E6SSM and F-NMSSM+ models discussed here the gauge coupling splitting 
due to flux in F-theory plays a crucial role.
The three different F-theory models are compared in Table \ref{models table}. 

In order for proton decay to be controlled, the geometric suppression at the field theory level corresponds to
the imposition of a discrete $Z_2$ symmetry. To understand the origin of this geometric suppression
would require knowledge of the GUT singlet matter curves, which in turn requires a knowledge of the global geometry. 
From our limited understanding of the global aspects of F-theory  
this just corresponds to an assumption about the global completion of the model.

\section*{Acknowledgements}

JCC and SFK would like to thank George Leontaris and Graham Ross for very useful discussions,
and for carefully reading the manuscript. SFK acknowledges partial support 
from the STFC Consolidated ST/J000396/1 and EU ITN grants UNILHC 237920 and INVISIBLES 289442.

\end{document}